 \documentclass[preprint,12pt]{elsarticle}
\usepackage{amssymb,amsmath,latexsym}
\usepackage[dvips]{color}
\usepackage[headings]{fullpage}
\usepackage{epic,epsfig,graphicx}
\usepackage{enumerate}
\usepackage{multirow}
\usepackage{setspace}
\usepackage{keyval}
\usepackage{float}

\usepackage{amssymb}

 \newtheorem{thm}{Theorem}[section]

 \newtheorem{lemma}{Lemma}[section]

 \newtheorem{defn}{Definition}[section]

 \newtheorem{corollary}{Corollary}[section]

 \newtheorem{rem}{Remark}[section]

 \numberwithin{equation}{section}
\journal{Journal of ***}

\begin{document}
\begin{frontmatter}
\title{On a family of stochastic SVIR influenza epidemic models and maximum likelihood estimation}
\author{Divine Wanduku }
\address{Department of Mathematical Sciences,
Georgia Southern University, 65 Georgia Ave, Room 3042, Statesboro,
Georgia, 30460, U.S.A. E-mail:dwanduku@georgiasouthern.edu;wandukudivine@yahoo.com\footnote{Corresponding author. Tel: +14073009605.
} }
\begin{abstract}
 This study presents a family of stochastic models for the dynamics of influenza in a closed human population.  We consider treatment for the disease in the form of vaccination, and incorporate the periods of effectiveness of the vaccine and infectiousness for the individuals in the population.  Our model is a SVIR model, with trinomial transition probabilities, where all individuals who recover from the disease acquire permanent natural immunity against the strain of the disease. Special SVIR models in the family are presented, based on the structure of the probability of getting infection and vaccination at any instant. The methods of maximum likelihood, and expectation maximization are derived for the parameters of the chain. Moreover, estimators  for some epidemiological assessment parameters, such as the basic reproduction number are computed. Numerical simulation examples are presented for the model.
\end{abstract}

\begin{keyword}
 Influenza epidemics \sep chain-trinomial model\sep MLE method\sep EM algorithm\sep basic reproduction number
\end{keyword}
\end{frontmatter}
\section{Introduction\label{ch1.sec0}}
Influenza ranks amongst the top ten most important diseases in the USA. In fact, the CDC estimates that from 2010-2011 to 2013-2014, influenza-associated deaths in the United States ranged from a low of 12,000 (during 2011-2012) to a high of 56,000 (during 2012-2013) annually \cite{estimate}. Globally, WHO estimates that seasonal influenza results in approximately 290-650 thousand deaths each year from just influenza related respiratory diseases alone\cite{WHO,lu}. These statistics necessitate more investigation and understanding about the disease, in order to control or ameliorate the burdens of the disease.

For hundreds of years, influenza, more commonly know as the flu, has plagued mankind.  The flu is caused by a virus, and spreads from person to person through direct contact and through particulates in the air.  Symptoms include-  fever, cough, sore throat, runny or stuffy nose, muscle or body aches, headaches, fatigue, and even vomiting and diarrhea \cite{FS&C}.

%

There are several strains of the influenza virus, all causing similar symptoms with varying severity.  Influenza viruses are generally categorized into four subgroups: A, B, C and D. Human influenza A and B viruses are the most common and cause seasonal epidemics of disease almost every winter in the United States. The emergence of a new and very different influenza A virus to infect people can cause an influenza pandemic. For instance, the recent "swine flu" (officially called H1N1), and "bird flu" (H5N1). See \cite{types}.

An individual infected with the flu can spread it to others up to about 6 feet away through droplets from their mouth from coughing, sneezing, or merely talking.
Once an individual has been infected with the flu virus, it usually takes about 2 days incubation period before symptoms start to show.  From the onset of symptoms, it can take 7 to 10 days for an individual to recover.  In general, adults who are infected with the influenza virus are capable of spreading it to others a day before symptoms showup as well as throughout the duration of the illness (cf.\cite{how}).
%

In an attempt to slow the spread of seasonal influenza, the CDC recommends annual influenza vaccination for people 6 months of age and older. The most common vaccination is the seasonal flu  injection, standard-dose trivalent shot (IIV3), and there are several other options available. One setback is that the flu virus constantly mutates and vaccination against one strain may not protect against other strains.  Once vaccinated, it may take up to two weeks before the vaccine is completely effective. See \cite{vacctypes}.
%

Various types of mathematical compartmental models have been used to investigate the dynamics of infectious diseases of humans\cite{tuckwell, wanduku-biomath,peter,Wanduku-2017,wanduku-delay,etba} . For instance, the authors in \cite{etba} present a deterministic model for influenza. The author in \cite{wanduku-biomath} presents an epidemic dynamic model for malaria. The  authors in \cite{peter,Wanduku-2017,wanduku-delay} present stochastic epidemic dynamic models with white noise perturbations for influenza.

The compartmental mathematical epidemic dynamic models are largely classified
as SVIS, SVIRS, SIS, SIR, SIRS, SEIRS, and SEIR etc. epidemic dynamic models depending on the compartments of the disease states directly involved in the general disease dynamics \cite{tuckwell, wanduku-biomath,peter,Wanduku-2017,wanduku-delay,etba}.
For instance, SIR model, represents transitions from the susceptible state to the infectious, and then to the removed state.   An SEIR model adds the incubation period via an exposed class $E$ to the previous SIR model. SVIR models incorporate the vaccination class $V$, and represent the dynamics of diseases that grant permanent immunity after recovery from disease, for example, influenza, where naturally acquired immunity against a strain of the virus is conferred after bacterial infection \cite{tuckwell, lloyd, SEIR, SVIS}.

Stochastic epidemic dynamic models more realistically represent epidemic dynamic processes because they include the randomness which inevitably occurs during a disease outbreak, owing to the presence of constant random environmental fluctuations in the disease dynamics. For stochastic epidemic models, the state of the system over time, which usually represents the number of people  susceptible (S), vaccinated (V), infected but not infectious (E), infected and infectious (I), or removed with natural immunity (R) is a stochastic process. Furthermore, the stochastic process for the disease dynamics can be a continuous-time-continuous state, continuous-state-discrete-time, or discrete-time and discrete state stochastic process. The choice on which type of stochastic process to use to represent the dynamics of the disease depends on the finiteness or infiniteness of the epidemiological features of the disease that are represented mathematically.  Some examples of continuous-time-continuous state stochastic epidemiological studies include \cite{Wanduku-2017,wanduku-delay,wanduku-fundamental,peter}, and discrete-time-discrete-state models include \cite{tuckwell, SHIRS, lloyd,novelframework}.

In our study, we derive an SVIR discrete-time and discrete state stochastic chain model, wherein temporary artificial immunity through vaccination is considered, in addition to the lifelong naturally acquired immunity for influenza in a closed community of human beings.

Chain binomial models are an important class of Markov chain models, and they have been used to represent and study the infectious diseases epidemic over time \cite{tuckwell, greenwood, reedfrost, wrong,  gani,SHIRS}. This family of stochastic models represent the disease dynamics as a random process, where transition between some states such as from susceptible to infectious, or from vaccinated to susceptible etc. are characterized by binomial transition probabilities.

 Some pioneer chain binomial models for infectious diseases include-  the Greenwood and Reed-Frost models\cite{greenwood,reedfrost}. Many other authors such as \cite{tuckwell,SHIRS,novelframework} have extended and expanded on the ideas of Greenwood and Reed-Frost, creating more realistic epidemic dynamic models.  Greenwood\cite{greenwood} in the 1931 study presented a SIS chain binomial chain model for the spread of diseases in human populations, where a breakdown of the population into successive generations indexed by $t = 0, 1, 2,\dots $ is used. Furthermore, the susceptible individuals are infected in a given generation, and are only capable of infecting others only within that generation.  And after which, they remain infected, but isolated from subsequent disease spread. Greenwood assumed that a susceptible person meets only one infectious person at any instance, and interaction with the  infectious person leads to transmission of disease with probability $p$.

 Reed-Frost\cite{reedfrost} built upon most of the ideas of the Greenwood model, and presented a SIR epidemic model. Moreover, Reed-Frost made a more realistic assumption that a susceptible person at any instance $t$ has a chance to interact independently with any of the infectious individuals present at that instance. Furthermore, the probability of getting infection at any instance $t$ is given by $p(y)=1-(1-p)^{y}$, where $y$ is the number of infectious people present at the instance $t$.

 	The previous Greenwood and Reed-Frost models consider generations of infections, and the infectious individuals no longer participate in subsequent transmission of disease to susceptible persons. This assumption is suitable for generalizations of disease dynamics, where  the disease suddenly outbreaks in a given time generation, dies out, and reoccur in another time generation.
	
	Tuckwell and Williams \cite{tuckwell} utilized some ideas from Reed-Frost\cite{reedfrost}, and proposed a more realistic SIR epidemic dynamic model. In their model, infection takes place over discrete time intervals, and the infectious population at any instance $t$ comprises of all infectious persons at different ages $k = 0,1,2,\ldots,R-2 $ of their infectiousness over the constant duration of the infectious period $R$. The age $k$ represents how long an individual has been infectious since initial infection from the susceptible state.  In their model, the total infectious individuals at any instance participates in the disease transmission process, and it is assumed that at the end of the infectious period $R$, all infectious persons are treated, fully recovered, and join the removed class $R$.

Tuckwell and Williams \cite{tuckwell} also assume that the $i^{th}$ individual in the population at any instant $t$ encounter a fixed number of people (family, office mates, employees) denoted $n_{i}$, and also a random number of people $M_{i}(t)$, where the random variables $M_{i}(t)$ are mutually independent and independent on the state of the population. This assumption leads to a more realistic random process description for the total number of people the  $i^{th}$ individual encounters at any  instance $N_{i}(t) = n_{i} + M_{i}(t), t\geq 0$. Moreover, the probability of getting infection at any instance becomes dependent on the number of infectious people the  $i^{th}$ susceptible individual meets among the $N_{i}(t)$ people.

Using ideas from Tuckwell-Williams \cite{tuckwell}, we propose a susceptible-vaccinated-infected-removed (SVIR) epidemic discrete-time and discrete state family of chain binomial models for influenza. 
Treatment in the form of vaccination is available, where the vaccine confers temporary effective artificial immunity against the disease, but ultimately wanes over time.
In line with \cite{tuckwell}, the infectious population involves in the disease transmission process over all generations of the population over time, and over all ages of infectiousness. However, unlike \cite{tuckwell}, the vaccination class is incorporated, and the total infectious and vaccinated populations are characterized based on the ages of infection and vaccination since initial infection and vaccination, over the constant and finite infectious and effective vaccination periods, respectively.
Moreover, as far as we know, no other discrete-time, discrete state chain binomial study in the line of thinking of Tuckwell-Williams\cite{tuckwell} with multiple time delays, and with the vaccination class exists in the literature. This study provides a suitable and more realistic extension of the chain binomial models- Greenwood, Reed-frost and Tuckwell-William models above.

The rest of this paper is organized as follows:  in Section~\ref{ch2.sec1}, we present an adequate description of the SVIR human population over discrete time.  In Section~\ref{ch2.sec3.sec1}, we derive the transition probabilities of the SVIR  Markov chain family of models, and present special cases of the general SVIR markov chain model.  In Section~\ref{sec4.sec5}, we present the method of maximum likelihood estimation to find important parameters of the SVIR chain models.  In Section ~\ref{sec5}, we estimate some important epidemiological parameters for the influenza epidemic model.  Finally, in Section~\ref{ch7}, we present numerical examples of the SVIR influenza model, and make general concluding remarks.

\section{Description of the SVIR Influenza Epidemic Process}\label{ch2.sec1}
In this section, we describe and represent the influenza epidemic in the human population.  We present the procedure to discretize time, and decompose the human population into the different states of the disease involved in the influenza epidemic

We consider a human population of size $n$ living in a closed natural environment over a period of time consistent with the duration of an epidemic outbreak, where it can be assumed that human movement into and out of the population is negligible or nonexistent. Furthermore, it is assumed that an infectious disease such as a flu, and other similar respiratory infectious diseases etc. breaks out in the closed population, and the disease causes major suffering in the human being, but with no mortal consequences due to disease related causes. In addition, the population is relatively safe, and no natural deaths occur  during the duration of the epidemic outbreak.  Also, it is assumed no births occur during the duration of the epidemic or very strong measures are taken medically to protect all newborns, such that they can be ignored from the effective diseased population.

It is also assumed that the infectious disease has an artificial vaccine which provides temporary effective immunity against the infectious agent that lasts over a constant immunity period $T_{1}$. At the end of the immunity period $T_{1}$, the vaccinated person becomes susceptible again to the disease. The disease also confers natural immunity after recovery from the infection. The naturally acquired immunity also provides effective natural protection against the disease agent, that lasts over a period of time longer than the duration of the epidemic in the population. Unlike the temporary artificial immunity period $T_{1}$, it is assumed that the naturally acquired immunity period is constant infinite. It is also assumed that the infectious agent has strong infectious abilities, such that all susceptible individuals who contract the infectious agent exhibit symptoms of illness after a relatively small time interval, and become infectious to other susceptible persons. Thus, the incubation period of the disease is considered small, and about one time unit, and consequently, the exposure class can be ignored. 

It is further assumed that the infectious period of all individuals who have contracted the disease is constant and finite, and it is denoted $T_{2}$. At the end of that period $T_{2}$, it is assumed that the immune system of the infected individual has established sufficient natural immunity against the particular strain of the disease, which lasts longer than the overall duration of the epidemic, and as a result individuals with naturally acquired immunity are removed from the epidemic process.

From the above description,  the human population of size $n$ is  subdivided into four major human subclasses  namely:  Susceptible ($S$), vaccinated ($V$), infectious ($I$), and removed ($R$). The susceptible class ($S$) does not have the disease, but are vulnerable to infection from the infectious class ($I$). A portion of the susceptible individuals is vaccinated artificially against the disease, and become the vaccinated class ($V$).  The vaccinated class ($V$) can no longer contract the disease during the effective artificially acquired temporary immunity period $T_{1}$ of the vaccine. At the end of the period $T_{1}$, the vaccine wanes, and the vaccinated individuals become susceptible again to the disease. When the infectious class ($I$) recovers  from the disease at the end of  the constant infectious period of time $T_{2}$, the individuals acquire natural immunity against the disease that can be considered to be constant and infinite. The individuals with naturally acquired immunity form the removed class denoted $R$. It is further assumed that the naturally acquired immunity is very effective such that the removed class never becomes susceptible again to the disease over the overall duration of the disease.

\begin{figure}[H]
\begin{center}
\includegraphics[scale=0.8]{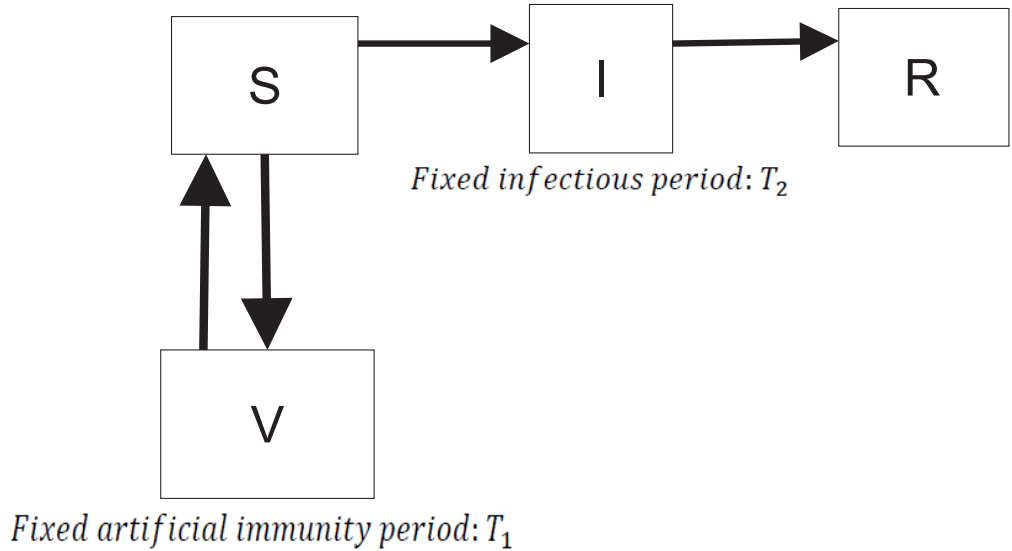}
\caption{Shows the framework of transitions between the different disease states (S,V,I,R) in the population during the outbreak of the disease.}\label{ch2.sec1.figure1}
\end{center}
\end{figure}

The general susceptible population $S(t)$ at any time $t=0, 1, 2, 3,\ldots$ over the duration of the epidemic, can be further broken down into two subclasses $\underbrace{S}(t)$ and $\overbrace{S}(t)$, where $\underbrace{S}(t)$ represents all susceptible individuals present at time $t$, who have never contracted the disease nor been vaccinated, and $\overbrace{S}(t)$ represents the number of susceptible individuals present at time $t$, who were previously vaccinated, and have lost the artificial immunity against the disease after the effective period $T_{1}$. That is,
\begin{equation}\label{ch2.sec1.eq1}
S(t)=\underbrace{S}(t)+\overbrace{S}(t).
\end{equation}
Note that over the duration of the epidemic, when it is assumed that all individuals are vaccinated only once against that strain of influenza virus, regardless whether the vaccine wanes over time, it follows that, only the class  $\underbrace{S}(t)$ will be liable to be vaccinated. When vaccination occurs multiple times, then all susceptible individuals $S(t)$ will be liable to be vaccinate at any time $t$. Also, at any time $t$,  the general susceptible class $S(t)$ is vulnerable to infection.
A compartmental framework exhibiting the transitions between the different states in the population is shown in Figure~\ref{ch2.sec1.figure1}

   In the following, we characterize the different disease subclasses namely:- susceptible, vaccinated, infectious and removed individuals over discrete time intervals of unit length, for example, days, weeks, months etc. The description  of the discretization of time is presented in the following.
   \begin{defn}\label{ch2.sec2.defn1.defn0} {Time discretization process:}

   The different disease subclasses namely:- susceptible, vaccinated, infectious and removal individuals shall be counted over discrete time intervals of unit length.
   That is, over the time subintervals $(t_0-1,t_{0}], (t_{0}, t_{0}+1], (t_{0}+1, t_{0}+2],(t_{0}+2, t_{0}+3],\ldots,(t_{0}+(t-2), t_{0}+(t-1)], (t_{0}+(t-1), t_{0}+t]$, where $t_{0}\geq 0$ is any nonnegative real number. Furthermore, the number of individuals in any subclass of the population at any time $t$, is taken to be the number of people in that state present in the subinterval of time $(t_{0}+(t-1), t_{0}+t]$, counted up to and including the point $t_{0}=t$,  where $t$ is a positive integer, that is, $t=0, 1,2,\ldots$. Moreover, all events that occur at the initial time $t=0$, refer explicitly to every event that occurred in the initial subinterval $(t_0-1, t_{0}]$, which forms an initial state for the epidemic. In addition, when $t_{0}=0$, then the initial interval $(t_0-1, t_{0}]$ reduces to the point $t=t_{0}=0$. 

    For example, the number of susceptible people present at time $t=2$, denoted $S(2)$, refers to the number of people in the susceptible state $S$, who are present in the subinterval of time $(t_{0}+1, t_{0}+2]$.  More generally, $S(t)$ is the number of susceptible people present in the subinterval $(t_0+(t-1),t_0+t]$.

Another discrete measure of time denoted by $k=0,1, 2,\ldots, t$, represents how long individuals in the different disease states $S, V, I$ and $R$, have been in the given states since their initial conversion into the states. We let $S_{k}(t), V_{k}(t), I_{k}(t) $ and $R_{k}(t),\forall k=0,1,2,3,\ldots, t$ denote the number of susceptible, vaccinated, infectious and removal individuals, respectively, present at time $t$, who have been in their different states for $k$ discrete time units, where $k=0,1,2,3,\ldots, t$. Note that according to the description of the time measure $k$, then $S_{0}(t), V_{0}(t), I_{0}(t) $ and $R_{0}(t)$ signify the number of people present at time $t$, who have just become susceptible, vaccinated, infectious and removed at time $t$.
\end{defn}
The decomposition of the susceptible population is presented in the following.
  \begin{defn}\label{ch2.sec2.defn1} {Decomposition and Aggregation of the Susceptible Population:}

  We decompose the susceptible population over the discrete times $t=0, 1, 2,..\cdots$  into subcategories based on (1) when a person first becomes susceptible to the disease, given that the person was previously vaccinated against the disease, and the artificial vaccine  has worn out after the effective vaccination period $T_{1}$, and (2) how long a person has been susceptible to the disease, since the initial time  $t=0$.

  For $t=0,1,2,\ldots$, let $S_{k}(t), \forall k=0,1,2,3,\ldots, t$ denote the number of susceptible individuals present at time $t$, who have been susceptible for $k$ discrete time units, where $k=0,1,2,3,\ldots, t$.  That is, the number of susceptible individuals $S(t)$ present at time $t=0,1,2,\ldots$ can be written in the two equivalent ways
  \begin{eqnarray}\label{ch2.sec2.def1.eq1}
   S(t)&=&S_{0}(t)+ S_{1}(t)+S_{2}(t)+S_{3}(t)+S_{4}(t)+\cdots \nonumber \\
   &&+ S_{t-3}(t)+S_{t-2}(t)+S_{t-1}(t)+S_{t}(t),\label{ch2.sec2.defn2.eq1}
   \end{eqnarray}
   where from (\ref{ch2.sec1.eq1}),
   \begin{eqnarray}
   \underbrace{S}(t)&=&S_{0}(t)+ S_{1}(t)+S_{2}(t)+S_{3}(t)+S_{4}(t)+\cdots \nonumber \\
   &&+ S_{t-3}(t)+S_{t-2}(t)+S_{t-1}(t),\label{ch2.sec2.defn2.eq1.eq1}
   \end{eqnarray}
   and
   \begin{equation}\label{ch2.sec2.defn2.eq1.eq2}
   \overbrace{S}(t)=S_{t}(t).
   \end{equation}
   From (\ref{ch2.sec2.defn2.eq1.eq1})-(\ref{ch2.sec2.defn2.eq1.eq2}), since infection and vaccination occur over every time $t=0,1,2,\ldots$, it follows that 
   $S_{k}(t)\leq S_{0}(t-k),\forall k=0,1,2,\ldots,t$, and  $S_{k}(k)\geq S_{k+1}(k+1),\forall k=0,1,2,\ldots,t-1$.

Furthermore, for any time $t$ at least as large as the maximum of the artificial immunity and the infectious periods $T_{1}$ and $T_{2}$, respectively, that is, for $t\geq max{(T_{1}, T_{2})}$
  it is easy to see that the relationship between susceptibility and vaccination is given as follows:
  \begin{eqnarray}\label{ch2.sec2.defn2.eq2}
  S_{k}(t)\leq S_{0}(t-k)=V_{T_{1}}(t-k)=V_{T_{1}+k}(t)=V_{0}(t-(T_{1}+k))\geq 0, k=0,1,\ldots,t-T_{1}.
    \end{eqnarray}

    It follows from (\ref{ch2.sec2.def1.eq1})-(\ref{ch2.sec2.defn2.eq2}) that the total susceptible population $S(t)=\underbrace{S}(t)+\overbrace{S}(t)$ at any time $t\geq T_{1}$, or $t\geq \max{(T_{1}, T_{2})}$, can be written as follows:
    \begin{equation}\label{ch2.sec2.defn2.eq4}
    \overbrace{S}(t) = S_{0}(t) + S_{1}(t) + S_{2}(t) + \cdots + S_{t-T_{1}}(t)	\quad and\quad \underbrace{S}(t)=S_{t}(t).
    \end{equation}
    Also, for any time $t< T_1$, or $t\leq \min{(T_{1}, T_{2})}$ then
    \begin{equation}\label{ch2.sec2.defn2.eq4.eq1}
    \overbrace{S(t)} =0,\quad and\quad \underbrace{S(t)} =  S_t(t).
    \end{equation}

       Also, under the assumption that the population is closed and there is no immigration of susceptible, vaccinated, infectious and removal individuals, then it follows from above that     $S(0)>0$, $S_{0}(0)\geq 0$, and $S_{k}(0)=0,\forall k\geq 1$.
  \end{defn}
  The decomposition of the vaccinated class $V(t)$ at time $t$ is presented in the following.
   \begin{defn}\label{ch2.sec2.defn3} {Decomposition and Aggregation of the Vaccinated Population:}

   We also decompose the vaccinated population class into subcategories based on how long individuals have been vaccinated before returning to the susceptible class of the population, whenever the artificial immunity wanes.  As with the previous definition, we decompose the vaccinated population over the discrete times $t=0, 1, 2,..\cdots$  into subcategories based on (1) when a person first becomes vaccinated against the disease, and (2) how long a person has been vaccinated against the disease, since the initial time  $t=0$, or simply how long a person has been vaccinated before the artificial immunity wears off completely after the period of effectiveness of the vaccines $T_{1}$. For $t=0,1,2,\ldots$, let $V_{k}(t), \forall k=0,1,2,3,\ldots, t$ denote the number of vaccinated individuals present at time $t$, who have been vaccinated for $k$ discrete time units, where $k=0,1,2,3,\ldots, t$.
   Letting $V(t)$ be the number vaccinated people present at time t, then it can be seen that for any $t=0,1,2,\dots$, then V(t) can be written in two different equivalent ways as follows:
For $t<T_{1}$ or $t<\min{(T_{1}, T_{2})}$,
\begin{equation}\label{ch2.sec2.def3.eq1}
   V(t)=V_{0}(t)+ V_{1}(t)+V_{2}(t)+V_{3}(t)+V_{4}(t)+\cdots+ V_{t-3}(t)+V_{t-2}(t)+V_{t-1}(t)+V_{t}(t)
   \end{equation}
   and since $V_{0}(t-k)=V_{k}(t), k=0,1,\ldots, t$, then $V(t)$ is equivalently expressed as follows:
   \begin{eqnarray}
   V(t)&=&V_{0}(t)+ V_{0}(t-1)+V_{0}(t-2)+V_{0}(t-3)\nonumber\\
   &&+V_{0}(t-4)+\cdots+ V_{0}(3)+V_{0}(2)+V_{0}(1)+V_{0}(0).\label{ch2.sec2.defn2.eq1}
  \end{eqnarray}

  Also for any time $t$ at least as large as $T_{1}$, or at least as large as the maximum of the artificial immunity and the infectious periods $T_{1}$ and $T_{2}$, respectively, that is, for $t\geq T_{1}$, or  $t\geq max{(T_{1}, T_{2})}$
  it is easy to see that some of the subclasses $V_k(t)$'s in  in the expression (\ref{ch2.sec2.def3.eq1}) are no longer in the vaccinated state, but have been converted into susceptible individuals.  That is,
  \begin{eqnarray}\label{ch2.sec2.def3.eq2}
   V_0(t-T_1)&=& V_{T_{1}}(t) = S_{0}(t), \nonumber \\
   V_0(t-(T_1+k))&=&V_{T_{1}+k}(t) =V_{T_{1}}(t-k)=S_{0}(t-k)\geq S_{k}(t), \forall k=0,1,\ldots,t-T_{1}\nonumber \\
    \end{eqnarray}

Therefore,  for $t\geq T_{1}$, or  $t\geq max{(T_{1}, T_{2})}$, the number of people in the vaccinated class at time $t$, is given by
\begin{eqnarray}\label{ch2.sec2.def3.eq3}
V(t)&=&V_0(t)+V_1(t)+V_2(t)+\dots +V_{T_1-1}(t)\nonumber\\
\textrm{or equivalently} \nonumber \\
V(t)&=&V_0(t)+V_0(t-1)+V_0(t-2)+ \dots + V_0(t-(T_1-1)).\nonumber\\
\end{eqnarray}
 \end{defn}

 The decomposition of the infectious class $I(t)$ at time $t$ is presented in the following.
   \begin{defn}\label{ch2.sec2.def4} {Decomposition and Aggregation of the Infectious Population:}

   We also decompose the infectious population class into subcategories based on how long individuals have been infectious before transitioning into the removed class of the population.  As with the previous definition, we decompose the infectious population over the discrete times $t=0, 1, 2,\ldots$  into subcategories based on (1) when a person first becomes infectious and (2) how long a person has been infectious, since the initial time  $t=0$, and until the infectious period $T_2$ is lapsed.

   For $t=0,1,2,\ldots$, let $I_{k}(t), \forall k=0,1,2,3,\ldots, t$ denote the number of infectious individuals present at time $t$, who have been infectious for $k$ discrete time units, where $k=0,1,2,3,\ldots, t$.  It follows that for any $t=0,1,2,\dots$, the number of infectious persons present at time $t$, denoted $I(t)$ is written as follows:
   For For $t<T_{2}$ or $t<\min{(T_{1}, T_{2})}$,
 \begin{eqnarray}\label{ch2.sec2.def4.eq1}
   I(t)&=&I_{0}(t)+ I_{1}(t)+I_{2}(t)+I_{3}(t)+I_{4}(t)+\cdots+ I_{t-3}(t)+I_{t-2}(t)+I_{t-1}(t)+I_{t}(t),\nonumber \\\textrm{or} \\
   I(t)&=&I_{0}(t)+ I_{0}(t-1)+I_{0}(t-2)+I_{0}(t-3)+I_{0}(t-4)+\cdots\nonumber\\
   &&+ I_{0}(3)+I_{0}(2)+I_{0}(1)+I_{0}(0),\label{ch2.sec2.defn3.eq1}
 \end{eqnarray}
 where $I_{k}(t)=I_{0}(t-k),\forall k=0,1,\ldots t$.

 In addition, for any time $t$ at least as large as the  infectious period $T_{2}$, or  at least as large as the maximum of the artificial immunity and the infectious periods $T_{1}$ and $T_{2}$ respectively, that is, for $t\geq T_{2}$, or  $t\geq max{(T_{1}, T_{2})}$,
  it is easy to see that some of the subcategories in the expression (\ref{ch2.sec2.def4.eq1}) are no longer in the infectious class, but have been converted into the removed state.  That is, for $t\geq T_{2}$, or  $t\geq max{(T_{1}, T_{2})}$, it is easy to see that
  \begin{equation}\label{ch2.sec2.defn3.eq1.eq1}
I(t)=I_0(t)+I_1(t)+I_2(t)+\dots +I_{T_2-1}(t),\nonumber\\
\end{equation}
or \\
\begin{equation}\label{ch2.sec2.defn3.eq1.eq2}
I(t)=I_0(t)+I_0(t-1)+I_0(t-2)+ \dots +  I_0(t-(T_2-1)),\nonumber\\
\end{equation}
and the following infectious classes are removed:
\begin{eqnarray}\label{ch2.sec2.def4.eq2.1}
  I_0(t-T_1)&=& I_{T_{2}}(t)=R_{0}(t),\nonumber \\
  I_0(t-(T_2+k))&=&I_{T_{2}+k}(t)=I_{T_{2}}(t-k)=R_{0}(t-k)=R_{k}(t),\forall k=0,1,2,\ldots, t-T_{2} \nonumber \\
  \end{eqnarray}
 \end{defn}

  The decomposition of the removed class $R(t)$ at time $t$ is presented in the following:
   \begin{defn}\label{ch2.sec2.def5} {Decomposition and Aggregation of the Removed Population:}

   Finally, we decompose the removed population class into subcategories based on how long individuals have been removed from the infectious state.  As with the previous definition, we decompose the removed population over the discrete times $t=0, 1, 2,\cdots$  into subcategories based on (1) when a person first becomes removed, and (2) how long a person has been removed from the infectious state since the initial time  $t=0$ of disease outbreak, or simply how long the person has been removed from the time when the person transitioned from the infectious state to the removal state. For $t=0,1,2,\ldots$, let $R_{k}(t), \forall k=0,1,2,3,\ldots, t$ denote the number of removed individuals present at time $t$, who have been removed for $k$ discrete time units, where $k=0,1,2,3,\ldots, t$. It follows that for any time $t=0,1,2,\dots$, the number of removed individuals at time $t$ denoted $R(t)$, can be written in two different equivalent ways as follows:
For $t<T_{2}$ or $t<\min{(T_{1}, T_{2})}$,
\begin{eqnarray}\label{ch2.sec2.def5.eq1}
     R(t) = R_t(t)=R_{0}(0)=R(0)\geq 0.
\end{eqnarray}
That is, $R(t)$ in (\ref{ch2.sec2.def5.eq1}) is the removed population at the initial time.

Furthermore, for any time $t$ at least as large as the infectious period $T_{2}$, or at least as large as the maximum of the artificial immunity and the infectious periods $T_{1}$ and $T_{2}$ respectively, that is, for $t\geq T_{2}$ or $t\geq max{(T_{1}, T_{2})}$,
 \begin{eqnarray}\label{2.2.5.1}
   R(t)&=&R_{0}(t)+ R_{1}(t)+R_{2}(t)+R_{3}(t)+R_{4}(t)+\cdots+ R_{t-3}(t)+R_{t-2}(t)+\nonumber \\
   &&R_{t-1}(t)+R_{t}(t) \nonumber\\
   \textrm{  or equivalently,} \nonumber \\
   R(t)&=&R_{0}(t)+ R_{0}(t-1)+R_{0}(t-2)+R_{0}(t-3)+R_{0}(t-4)+\cdots+ \nonumber \\
   &&R_{0}(3)+R_{0}(2)+R_{0}(1)+R_{0}(0),
 \end{eqnarray}
where $R_{k}(t)=R_{0}(t-k),\forall k=0,1,\ldots t$.

It is also easy to see that some of the subcategories in (\ref{2.2.5.1}) are equivalent to some infectious subclasses as shown below:
\begin{eqnarray}\label{ch2.sec2.defn5.eq2}
  R_{0}(t)&=&I_{T_{2}}(t)=I_{0}(t-T_{2}), \nonumber \\
 R_{k}(t)&=&R_0(t-k)=I_{T_{2}}(t-k)=I_{T_{2}+k}(t)=I_{0}(t-(T_{2}+k)),k=0,1,\ldots, t-T_{2}. \nonumber\\
     \end{eqnarray}

Therefore, for $t\geq T_{2}$ or $t\geq max{(T_{1}, T_{2})}$, the number of removed individuals at time $t$, $R(t)$ is given by:
\begin{eqnarray}\label{ch2.sec2.defn5.eq3}
R(t)&=&R_0(t)+R_1(t)+R_2(t)+\dots +R_{t-T_2}(t)\nonumber\\
\textrm{ or}\nonumber \\
R(t)&=&R_0(t)+R_0(t-1)+R_0(t-2)+ \dots + R_0(t-(t-T_2)).\nonumber\\
&=&R_0(t)+R_0(t-1)+R_0(t-2)+ \dots + R_0(T_2).
\end{eqnarray}
\end{defn}
\begin{defn}\label{ch2.sec2.defn6} {Decomposition and Aggregation of the total human population:}

It can be seen from Definitions \ref{ch2.sec2.defn1} - \ref{ch2.sec2.def5} that the total human population of size $n$ can be written as follows:\\
For all $t \in \mathbb{Z}_+$ = {0,1,2,3,...}
\begin{eqnarray}\label{ch2.sec2.def6.eq1}
n&=& S(t)+V(t)+I(t)+R(t), \\
\text{or} \nonumber \\
n&=& \underbrace{S(t)} + \overbrace{S(t)} + V(t) + I(t) + R(t). \label{ch2.sec2.def6.eq1a}
\end{eqnarray}
So for $T_1 \leq t < T_2$, the vector
\[B(t) = (S(t)=\underbrace{S(t)}+\overbrace{S(t)}, V_0(t), V_1(t), \dots, V_{T_1-1}(t), I_0(t), I_1(t), \dots, I_t(t))\] is sufficient to define all states of the process.  Also, \\
\begin{eqnarray}
	n=\underbrace{S(t)} + \underset{i=0}{\overset{t-{{T}_{1}}}{\mathop \sum }}\,{{S}_{i}}\left( t \right)+ \underset{i=0}{\overset{{{T}_{1}}-1}{\mathop \sum }}\,{{V}_{i}}\left( t \right) + \underset{i=0}{\overset{t}{\mathop \sum }}\,{{I}_{i}}\left( t \right)+ R(t), \label{ch2.sec2.def6.eq3}
\end{eqnarray}
where $R_t(t)=R(0) \geq 0$.

For $T_2 \leq t < T_1$, the vector
\[B(t)=(S(t)=\underbrace{S(t)},V_0(t), V_1(t), \dots, V_t(t), I_0(t), I_1(t), \dots, I_{T_2-1}(t))\]\\
is sufficient to define all states of the process.  Also, \\
\begin{eqnarray}
	n=\underbrace{S(t)} + \underset{i=0}{\overset{t}{\mathop \sum }}\,{{V}_{i}}\left( t \right) + \underset{i=0}{\overset{{{T}_{2}}-1}{\mathop \sum }}\,{{I}_{i}}\left( t \right) + \underset{i=0}{\overset{t-{{T}_{2}}}{\mathop \sum }}\,{{R}_{i}}\left( t \right).\label{ch2.sec2.def6.eq3}
\end{eqnarray}
Also, for  $t\leq \min{(T_{1}, T_{2})}$, the vector
\[B(t)=(S(t), V_0(t), V_1(t), \dots, V_t(t), I_0(t), I_1(t), \dots, I_t(t))\] is sufficient to define all states of the process.  Also, \\
\begin{eqnarray}\label{ch2.sec2.def6.eq2}
n= \underset{i=0}{\overset{t}{\mathop \sum }}S_i(t) + \underset{i=0}{\overset{t}{\mathop \sum }}V_i(t) + \underset{i=0}{\overset{t}{\mathop \sum }}I_i(t) + R_t(t),
\end{eqnarray}
where $R_t(t)=R(0) = 0$.

In addition, for $t\geq max{(T_{1}, T_{2})}$, the vector \\
\[B(t)=(S(t)=\underbrace{S(t)}+\overbrace{S(t)}, V_0(t), V_1(t), \dots, V_{T_1-1}, I_0(t), I_1(t), \dots, I_{T_2-1}(t))\] is sufficient to define all states of the process.  Also, \\
\begin{eqnarray}
	n=\underbrace{S(t)} + \underset{i=0}{\overset{t-{{T}_{1}}}{\mathop \sum }}\,{{S}_{i}}\left( t \right)+ \underset{i=0}{\overset{{{T}_{1}}-1}{\mathop \sum }}\,{{V}_{i}}\left( t \right) + \underset{i=0}{\overset{{{T}_{2}}-1}{\mathop \sum }}\,{{I}_{i}}\left( t \right) + \underset{i=0}{\overset{t-{{T}_{2}}}{\mathop \sum }}\,{{R}_{i}}\left( t \right).\label{ch2.sec2.def6.eq3}
\end{eqnarray}
\end{defn}
\section{Derivation of the SVIR Influenza Markov Chain Model and Transition Probabilities}\label{ch2.sec3.sec1}
In this section we derive the SVIR chain-binomial epidemic dynamic model for the influenza epidemic.  The epidemic model presented is an extension of the studies by \cite{tuckwell,greenwood,marco}. That is, we derive a random process that characterizes the dynamics of influenza in the human population  structured over time as described in above, and  derive the general formula for the transition probabilities of the Markov chain.
Define the following non-negative integer valued random vector function:
\begin{equation}\label{ch2.sec3.eq5}
B: \mathbb{Z} _{+} \longrightarrow \mathbb{Z} _{+}^{T_1+T_2+1},\end{equation}
where for each $t\in \mathbb{Z} _{+}$, and $t\geq max\{T_1, T_2\}$,  the random vector $B$ is given by
\begin{equation}\label{3.0A}
       B(t) = (S(t), V_0(t), V_1(t), \dots, V_{T_1-1}(t), I_0(t), I_1(t), \dots, I_{T_2-1}(t)) \in \mathbb{Z}_{+}^{T_1+T_2+1}.
       \end{equation}
     Also, for each $t\in \mathbb{Z} _{+}$,  an observed vector value $b(t)$ for the random vector $B(t)$ is defined as follows:
     \begin{equation}\label{ch2.sec3.eq6a}
          b(t)=(x_t, y_{0_t}, y_{1_t}, y_{2_t}, \dots, y_{(T_1-1)_t}, z_{0_t}, z_{1_t}, \dots, z_{(T_2-1)_t}) \in \mathbb{Z}_+^{T_1+T_2+1}.
     \end{equation}
     Note that for each $t\in \mathbb{Z} _{+}$, the observed values $x_t, y_{0_t}, y_{1_t}, y_{2_t}, \dots, y_{(T_1-1)_t}, z_{0_t}, z_{1_t}, \dots, z_{(T_2-1)_t}$ are non-negative integers.
             Furthermore,
        \[
    B(t) = b(t)
    \]
    if and only if
    \begin{eqnarray}
        &&S(t) = x_t,\quad V_0(t)= y_{0_t},\quad V_1(t)=y_{1_t}, \dots, V_{T_1-1}(t)= y_{(T_1-1)_t},\nonumber\\
        && I_0(t)=z_{0_t}, \dots, I_{T_2-1}(t)=z_{(T_2-1)_t}.\label{ch2.sec3.eq7}
    \end{eqnarray}
The random process $\{B(t), t=1,2,3,\dots\}$, describes the evolution of the influenza epidemic in the $SVIR$ population characterized above.

Furthermore, for $t<min\{T_1, T_2\},$
\begin{equation}\label{ch2.sec3.eq7.eq1}
    B(t)=(S(t), V_0(t), V_1(t),V_2(t). \dots, V_t(t), I_0(t), I_1(t), \dots, I_t(t))
\end{equation}
is sufficient to define all the states of the process.
 For $T_1 \leq t < T_2$
\begin{equation}\label{ch2.sec3.eq7.eq2}
    B(t)=(S(t)=\underbrace{S(t)}+\overbrace{S(t)}, V_0(t), V_1(t), \dots, V_{T_1-1}(t), I_0(t), I_1(t), \dots, I_t(t))
\end{equation}
is sufficient to define all the states of the process.
 For $T_2 \leq t < T_1$
\begin{equation}\label{ch2.sec3.eq7.eq3}
    B(t)=(S(t)=\underbrace{S(t)}, V_0(t), V_1(t), \dots, V_t(t), I_0(t), I_1(t), \dots, I_{T_2-1}(t))
\end{equation}
is sufficient to define all states of the process.
And for $t\geq max\{T_1, T_2\}$
\begin{equation}\label{ch2.sec3.eq7.eq4}
    B(t)=(S(t)=\underbrace{S(t)}+\overbrace{S(t)}, V_0(t), V_1(t), \dots, V_{T_1-1}(t), I_0(t), I_1(t), \dots, I_{T_2-1}(t))
\end{equation}
is sufficient to define all states of the process.

It should be noted from (\ref{ch2.sec3.eq7.eq1}) - (\ref{ch2.sec3.eq7.eq4}) that the random process $\{B(t): t\geq 0\}$  takes different simplifications for different sub-intervals of the time interval, $t \in [0,\infty)$.  Without loss of generality, we present the general form of the transition probability of the random process $\{B(t): t \geq 0\}$ in the case where $t \geq max \{T_1, T_2\}$.
In the following theorem, we show that the random process $\{B(t), t=1,2,3,\dots\}$ is a Markov chain.
\begin{thm}\label{ch2.sec3.thm1}
For $t\geq max\{T_1, T_2\}$, the random process $\{B(t), t=1,2,3,\dots\}$, where the $(T_1+T_2+1)$ non-negative integer value random vector function $B(t)=(S(t)
V_0(t), V_1(t), \dots, V_{T_2-1}(t),
I_0(t), I_1(t), \dots, I_{T_2-1}(t)),\forall t=0,1,\ldots$, defines a Markov chain for all $t=0,1,2,3,...$.  That is,
   Furthermore, the transition probabilities are completely specified by the conditional distribution of the states $V_{0}$ and $I_{0}$. That is,
   \begin{equation}\label{ch2.sec3.thm1.eq2}
   P(B(t+1) = b(t+1) \vert B(t)= b(t))=P(V_0(t+1)=y_{0_{t+1}}, I_0(t+1)=z_{0_{t+1}}\vert B(t)=b(t)),
   \end{equation}
   where the terms $S(t+1)=x_{t+1}$, $V_{0}(t+1)=y_{0_{t+1}}$, $I_{0}(t+1)=z_{0_{t+1}}$, and $S(t)=x_{t}$ are related as follows
   \begin{equation}\label{ch2.sec3.thm1.eq3}
   y_{0_{t+1}} = x_t-x_{t+1}-z_{0_{t+1}}.
   \end{equation}
   That is,
   \item[1.] If $S(t+1)=x_{t+1}=0$, then
   \begin{eqnarray}
   P(B(t+1) = b(t+1) \vert B(t)= b(t))&=&P(V_0(t+1)=y_{0_{t+1}}, I_0(t+1)=x_{t}-y_{0_{t+1}}\vert B(t)=b(t)),\nonumber\\
   &=&P(V_0(t+1)=x_{t}-z_{0_{t+1}}, I_0(t+1)=z_{0_{t+1}}\vert B(t)=b(t)),\nonumber\\\label{ch2.sec3.thm1.eq4}
   \end{eqnarray}
   \item[2.] If $S(t+1)=x_{t+1}=x_{t}$, then
   \begin{eqnarray}
   P(B(t+1) = b(t+1) \vert B(t)= b(t))&=&P(V_0(t+1)=0, I_0(t+1)=0\vert B(t)=b(t)),\nonumber\\\label{ch2.sec3.thm1.eq5}
   \end{eqnarray}
   \item[3.] If $S(t+1)=x_{t+1}\in (0,x_{t})$, then
   \begin{eqnarray}
   P(B(t+1) = b(t+1) \vert B(t)= b(t))&=&P(V_0(t+1)=x_{t}-x_{t+1}-z_{0_{t+1}}, I_0(t+1)=z_{0_{t+1}}\vert B(t)=b(t)),\nonumber\\
   &=&P(V_0(t+1)=y_{0_{t+1}}, I_0(t+1)=x_{t}-x_{t+1}-z_{0_{t+1}}\vert B(t)=b(t)).\nonumber\\\label{ch2.sec3.thm1.eq6}
   \end{eqnarray}
\end{thm}
Proof:\\
As mentioned earlier, we show that $B(t)$ is a Markov chain without loss of generality for $t \geq \max{(T_1, T_2)}$.  Moreover, we assume further without loss of generality that $T_1 \geq T_2$. Thus, $max(T_1, T_2)=T_1$. To show that $\{B(t), t\in \mathbb{Z}_+\}$ is a Markov chain, it suffices to show that the Markov property holds in $\{{B(t), t\in \mathbb{Z}_+}$\}.  That is, we show that for all t $\geq max(T_1, T_2) = T_1 \geq T_2,$
\begin{eqnarray}\label{ch2.sec3.prf1.eq1}
&&P(B(t+1)=b(t+1) \vert B(t) = b(t), B(t-1)=b(t-1), \dots, B(1)=b(1), B(0)=b(0)) \nonumber \\
&&= P(B(t+1)=b(t+1) \vert B(t)=b(t)),
\end{eqnarray}
where $B(t)$ and $b(t)$ are defined in (\ref{3.0A}) and (\ref{ch2.sec3.eq6a}).
Indeed, from (\ref{ch2.sec3.eq7.eq4}),
\begin{eqnarray}
P(B(t+1)&=&b(t+1)\vert B(t)=b(t), B(t-1)=b(t-1), \dots, B(1)=b(1), B(0)=b(0)) \nonumber\\
&=&P({S(t+1)}=x_{(t+1)}, V_0(t+1)=y_{0_{t+1}}, \dots, V_{(T_1-1)}(t+1)=y_{(T_1-1)_{t+1}},\nonumber\\
&& I_0(t+1)=z_{0_{t+1}}, \dots, I_{(T_2-1)}(t+1)=z_{(T_2-1)_{(t+1)}}\vert B(t)=b(t), \dots, B(0)=b(0))\nonumber\\
&\equiv&RHS.\label{ch2.sec3.prf1.eq2}
\end{eqnarray}
Observe that at time $t+1$, $S(t+1)$ is expressed as follows:
\begin{equation}\label{ch2.sec3.prf1.eq4}
{S(t+1)} = {S(t)} -V_{0}(t+1) - I_{0}(t+1), \quad \forall  t  \geq max{(T_{1}, T_{2})}.
\end{equation}
The right-hand-side (RHS) of (\ref{ch2.sec3.prf1.eq2}) can be expressed explicitly as follows:
\begin{eqnarray}
RHS\equiv P({S_{t+1}(t+1)} = x_{t+1} , V_{0}(t+1) = y_{0_{t+1}}, V_{1}(t+1)=y_{1_{t+1}}, V_{2}(t+1) = y_{2_{t+1}}, \dots \nonumber \\
V_{T_{1}-1}(t+1) = y_{T_{1}-1_{t+1}}, I_{0}(t+1)=z_{0_{t+1}}, I_{1}(t+1) = z_{1_{t+1}}, I_{2}(t+1)=z_{2_{t+1}}, \dots \nonumber \\
I_{t_{2} -1}(t+1)= z_{t_{2}-1_{t+1}}\vert \underbrace{S_{t}(t)} = x_{t} , V_{0}(t) = y_{0_{t}}, V_{1}(t)=y_{1_{t}}, V_{2}(t) = y_{2_{t}}, \dots \nonumber \\ V_{T_{1}-1}(t) = y_{T_{1}-1_{t}}, I_{0}(t)=z_{0_{t}}, I_{1}(t) = z_{1_{t}}, I_{2}(t)=z_{2_{t}}, \dots, I_{t_{2} -1}(t)= z_{t_{2}-1_{t}}, \nonumber \\B(t-1)=b(t-1), B(t-2)=b(t-2), \dots, B(1)=b(1), B(0)=b(0)).\nonumber\\
\label{ch2.sec3.prf1.eq6}
\end{eqnarray}
By applying (\ref{ch2.sec3.prf1.eq4}) to (\ref{ch2.sec3.prf1.eq6}),
it is easy to see that (\ref{ch2.sec3.prf1.eq6}) reduces to
\begin{eqnarray}
&&RHS\nonumber\\
&&\equiv P(V_{0}(t+1) = y_{0_{t+1}} = x_{t} - x_{t+1} - z_{0_{t+1}}, V_{1}(t+1)=y_{0_{t}}, V_{2}(t+1) = y_{1_{t}}, \dots \nonumber \\ &&V_{(T_{1}-1)}(t+1) = y_{(T_{1}-2)_{t}},
I_{0}(t+1)=z_{0_{t+1}}, I_{1}(t+1) = z_{0_{t}},\nonumber \\ &&I_{2}(t+1)=z_{1_{t}}, \dots, I_{(t_{2} -1)}(t+1)=z_{(t_{2}-2)_{t}}\vert S_{t}(t) = x_{t}, \nonumber \\
&& V_{0}(t) = y_{0_{t}}, V_{1}(t)=y_{1_{t}}, V_{2}(t) = y_{2_{t}}, \dots V_{(T_{1}-1)}(t) = y_{(T_{1}-1)_{t}}, \nonumber \\
&&I_{0}(t)=z_{0_{t}}, I_{1}(t) = z_{1_{t}}, I_{2}(t)=z_{2_{t}}, \dots,I_{t_{2} -1}(t)= z_{(t_{2}-1)_{t}},\nonumber \\
&&B(t-1)=b(t-1), B(t-2)=b(t-2), \dots, B(1)=b(1), B(0)=b(0)).\nonumber \\\label{ch2.sec3.prf1.eq7}
\end{eqnarray}
Since most of the terms in the argument of the conditional probability in (\ref{ch2.sec3.prf1.eq7}) are already given as the conditions of the probability, and $V_0(t+1)$ and $I_0(t+1)$ have no relationship with $B(t-1), B(t-2),\dots, B(1),$ and $B(0)$, the Markov property in (\ref{ch2.sec3.prf1.eq1}) follows immediately. Moreover, (\ref{ch2.sec3.prf1.eq7}) further reduces to:
\begin{eqnarray}
RHS&\equiv&P(V_{0}(t+1)=y_{0_{t+1}}=x_{t} - x_{t+1} - z_{0_{t+1}}, I_{0}(t+1) = z_{0_{t+1}}\vert {S(t)} = x_{t}, V_{0}(t) = y_{0_{t}},\nonumber \\
&&V_{1}(t) = y_{1_{t}}, \dots,
V_{T_{1}-1}(t) = y_{(T_{1}-1)_{t}}, I_{0}(t) = z_{0_{t}}, \nonumber \\
&&I_{1}(t) = z_{1_{t}}, \dots, I_{(T_{2}-1)}(t) = z_{(T_{2}-1)_{t}}),\label{ch2.sec3.prf1.eq8}
\end{eqnarray}
or to
\begin{eqnarray}
RHS&\equiv&P(V_{0}(t+1)=y_{0_{t+1}}, I_{0}(t+1) = z_{0_{t+1}}=x_{t} - x_{t+1} - y_{0_{t+1}}\vert  S(t) = x_{t}, V_{0}(t) = v_{0_{t}}, \nonumber \\
&&V_{1}(t) = y_{1_{t}}, \dots, V_{T_{1}-1}(t) = y_{(T_{1}-1)_{t}}, I_{0}(t) = z_{0_{t}}, \nonumber \\
&&I_{1}(t) = z_{1_{t}}, \dots, I_{(T_{2}-1)}(t) = z_{(T_{2}-1)_{t}}).\label{ch2.sec3.prf1.eq8a}
\end{eqnarray}
Thus, the second part (\ref{ch2.sec3.thm1.eq2}) is proved.
\begin{rem}\label{ch2.sec3.thm1.rem1}
In reality, the relationship between the events of getting vaccinated and getting infected exhibit various mathematical forms depending on the properties of the particular disease scenario. That is, the joint conditional distributions of the random variables $V_{0}(t)$  and $I_{0}(t)$, for any time $t=0,1,\ldots$ can be be expressed in various forms depending on the disease scenario.

For instance, in more organized societies, the decision to be vaccinated, in some cases, is not influenced by the outbreak of influenza, since influenza seasons are predictable and periodic vaccination against influenza is encouraged.  In other disease scenarios especially when the influenza outbreak is unpredictable,  more people tend to get vaccinated whenever the influenza epidemic breaks-out, and more influenza cases have been reported. Also, in other influenza scenarios, the vaccine may not be sufficiently strong to prevent infection, but mainly to reduce the severity of infection.  These scenarios can also create independence between vaccination and infection.  One of such special influenza disease scenarios is described in the  next section.

Observe Theorem~\ref{ch2.sec3.thm1}[1.] for $S(t+1)=x_{t+1}=0$ signifies that at the $t+1$ time step, all susceptible susceptible individuals $x_{t}$ are either vaccinated or infected by the virus.  Theorem~\ref{ch2.sec3.thm1}[2.] for $S(t+1)=x_{t+1}=x_{t}$ signifies that no infection or vaccination occurs at the   $t+1$ time step, and Theorem~\ref{ch2.sec3.thm1}[3.] for $S(t+1)=x_{t+1}\in (0,x_{t})$ signifies that either $y_{0_{t=1}}$ number of people are vaccinated, $x_{t}-x_{t+1}-y_{0_{t=1}}$ are infected, and $x_{t+1}$ remain susceptible at time $t+1$, or $z_{0_{t=1}}$ number of people are infected, $x_{t}-x_{t+1}-z_{0_{t=1}}$ are vaccinated, and $x_{t+1}$ remain susceptible at time $t+1$. Thus, the transition probabilities (\ref{ch2.sec3.thm1.eq4})-(\ref{ch2.sec3.thm1.eq6}) represent a tri-variate distributions distributions for $V_{0}$ and $I_{0}$.
\end{rem}
\section{Special SVIR Models}\label{sec4}
 As remarked in  Theorem~\ref{ch2.sec3.thm1}, the transition probabilities of the chain $\{B(t), t=0,1,2,\dots\}$ are completely defined by the joint conditional distribution of the discrete random variables $V_0$ and $I_0$.  There are several possible discrete tri-variate distributions that can be used to specify the transition probabilities for the chain. In this section, we consider two different cases based on the definition of the probability of getting vaccination, infection or remaining susceptible in the next time step, for each susceptible individual in the population. The trinomial distribution will be used to characterize the transition probabilities for the chain $\{B(t), t=0,1,2,\dots\}$.
%
%
\subsection{The SVIR Influenza Model with vaccination dependent on infectious encounters } \label{sec4.subsec1}
The following assumptions are made for the influenza epidemic in this subsection:- (a) it is assumed that the influenza epidemic is very severe and highly contagious, and the population is also well sensitized about the disease, and there are available vaccines for those seeking vaccination. Furthermore, because of the high prevalence of the disease, every susceptible person in the population at the end of any time interval $t$, that is, $(t-1,t]$ is either infected, vaccinated or remains susceptible. Moreover, the probability of  the $i^{th}$ susceptible individual from the $(t-1)^{th}$ generation $S(t-1)=x_{t-1}$ escaping infection, and avoiding vaccination at time $t$ is denoted $P^{i}_{S}(t)$.

(b) Using ideas from \cite{reedfrost,tuckwell}, we assume further that  $p $ is the probability of becoming infected after one interaction with an infected individual, and all interactions between a susceptible person and infectious individuals are independent. Thus, at time $t$, the probability that the $i^{th}$ susceptible individual from the $(t-1)^{th}$ generation $S(t-1)=x_{t-1}$ becomes infectious after interaction with $j$ infectious individuals is denoted $p_j^i(t) \equiv p_j^i$, and is given by
\begin{equation}\label{sec4.subsec1.eq1}
    p_j^i=1-(1-p)^j, i=1, 2, 3,\dots, x_{t-1}, \text{where} \quad S(t-1)=x_{t-1}.
\end{equation}

(c) Let $N_{i}(t)$ represent the number of people the $i^{th}$ susceptible individual interacts with at time $t$.  Note that $N_i(t)$ is a random variable for each $t=0,1,2,\dots$.  In fact, $\{N_i(t), t=0, 1, 2, \dots\}$ is a random process. For the purpose of illustrating the method of expectation-maximization (EM) algorithm to find the maximum likelihood estimators (MLE's) for the parameters of the influenza epidemic model, we shall assume that $N_i(t)$ is a constant $N>0$, that is $N_i(t)\equiv N$.  Moreover, we assume that $N$ is chosen in such a way that the binomial approximation for the number of infectious people a susceptible person meets can be used. In the next subsection, we characterize the random process $\{N_i(t), t=0, 1, 2, \dots\}$ as a poisson process.

(d) Furthermore, it is assumed that there is homogenous mixing in the population, and as a result, the probability that the $i^{th}$ susceptible person $S(t-1)=x_{t-1}$ at the beginning of the $t^{th}$ time interval $(t-1,t]$  meets an infectious person $I(t-1)$ also from the $(t-1)^{th}$ generation present at time $t$ is given as $\frac{I(t-1)}{n-1}$.  Thus, the probability that the $i^{th}$ individual meets exactly $j$ infectious individuals is defined as follows:
\begin{equation}\label{B6}
P_{j}^{i}(N_i(t)) = \binom{N_{i}(t)}{j}\left(\frac{I(t-1)}{n-1}\right)^{j}\left(1-\frac{I(t-1)}{n-1}\right)^{N_{i}(t)-j},
\end{equation}

It follows from the assumptions (a)-(d) above that the probability of the $i^{th}$ susceptible individual of $S(t-1)=x_{t-1}$ meeting infectious individuals and becoming infected at time $t$  is given by
\begin{equation}\label{B8}
P_I^i(t)=\sum_{j=1}^{N_i(t)}(1-(1-p)^j)\binom{N_i(t)}{j}\left(\frac{I(t-1)}{n-1}\right)^j\left(1-\frac{I(t-1)}{n-1}\right)^{N_i(t)-j},
\end{equation}
which simplifies to
\begin{equation}\label{4.2C}
    P_I^i(t)\equiv P_I^t=1-\left(1-\frac{pI(t-1)}{n-1}\right)^{N_i(t)}.
\end{equation}

(e) It is assumed further that susceptible individuals are motivated to seek vaccination due to their encounter with infectious individuals, and such encounters within any one time unit which do not lead to infection, give rise to immediate desire in the susceptible person to get vaccinated within the one time unit. Furthermore, it is assumed that the disease has an incubation period at most one time unit, and the vaccine is relatively strong to reverse any infection that occurs within the span of one time unit, such that all susceptible people who get infected and are also vaccinated within the one time unit, obtain artificial immunity, and join the vaccinated class. Suppose a susceptible person eludes infection within a one time unit, let $\phi\in (0,1)$ be the probability that the susceptible individual develops desire and gets vaccinated within the time interval.

Also, from assumption (a), since three outcomes are possible for any susceptible person at time $t$, therefore, the probability that the $i^{th}$ susceptible person $S(t-1)=x_{t-1}$ at the beginning of the $t^{th}$ time interval $(t-1,t]$  meets an infectious person $I(t-1)$ also from the $(t-1)^{th}$ generation present at time $t$, and does not become infected, and becomes vaccinated at time $t$, denoted $P^{i}_V(t)$, satisfies $P^{i}_V(t)+P^{i}_I(t)=1-P^{i}_{S}(t)$, and is derived as follows:
  \begin{equation}\label{4.2C.eq1}
P^{i}_V(t)=\sum_{j=1}^{N_i(t)}(\phi(1-p)^j)\binom{N_i(t)}{j}\left(\frac{I(t-1)}{n-1}\right)^j\left(1-\frac{I(t-1)}{n-1}\right)^{N_i(t)-j},
\end{equation}
which simplifies to
\begin{equation}\label{4.2C.eq2}
    P_V^i(t)\equiv P_V^t=\phi\left(1-\frac{pI(t-1)}{n-1}\right)^{N_i(t)}.
\end{equation}
we now present the transition probabilities for the SVIR stochastic process $\{B(t), t=1,2,3,\dots\}$, whenever $t\geq max\{T_1, T_2\}$.
\begin{thm}\label{sec4.subsec1.thm1}
Let the conditions of Theorem~\ref{ch2.sec3.thm1} be satisfied, and let the probability that the $i^{th}\forall i=1,2,\ldots,x_{t}$ susceptible gets vaccinated at time $t$, denoted $P^{i}_V(t)$ be as defined in (\ref{4.2C.eq2}), the probability that the $i^{th},\forall i=1,2,\ldots,x_{t}$ susceptible gets infected at time $t$, denoted $P^{i}_I(t)$ be  as defined in (\ref{4.2C}), where $P^{i}_V(t)+P^{i}_I(t)+P^{i}_S(t)=1$. It follows that for $t\geq max\{T_1, T_2\}$, the transition probabilities for the stochastic process $\{B(t), t=1,2,3,\dots\}$, are given as follows. For $S(t+1)=x_{t+1}\in (0,x_{t})$, and with only every susceptible individual vaccinated only once, it follows that
%
\begin{eqnarray}
    P(B(t+1)=b_{t+1}\vert B(t))&=&\binom{\underbrace{x_t}}{y_{0_{t+1}}}\binom{x_{t}-y_{0_{t+1}}}{z_{0_{t+1}}}\left(P^{i}_V(t)\right)^{y_{0_{t+1}}}
    \left(P^{i}_I(t)\right)^{z_{0_{t+1}}}\times\nonumber\\
    &&\times  \left(P^{i}_S(t)\right)^{x_{t}-z_{0_{t+1}}-y_{0_{t+1}}}
        \label{sec4.subsec1.thm1.eq1}
\end{eqnarray}
where $x_{t+1}+y_{0_{t+1}}+z_{0_{t+1}}=x_{t}$.

For $S(t+1)=x_{t+1}\in (0,x_{t})$, and with multiple vaccination, it follows that
%

\begin{eqnarray}
    P(B(t+1)=b_{t+1}\vert B(t))&=&\binom{{x_t}}{y_{0_{t+1}}}\binom{x_{t}-y_{0_{t+1}}}{z_{0_{t+1}}}\left(P^{i}_V(t)\right)^{y_{0_{t+1}}}
    \left(P^{i}_I(t)\right)^{z_{0_{t+1}}}\times\nonumber\\
    &&\times  \left(P^{i}_S(t)\right)^{x_{t}-z_{0_{t+1}}-y_{0_{t+1}}}
        \label{sec4.subsec1.thm1.eq1.eq1}
\end{eqnarray}
where $x_{t+1}+y_{0_{t+1}}+z_{0_{t+1}}=x_{t}$.
In addition, the conditional marginal distributions of $V_{0}$ and $I_{0}$ are given as follows:
\begin{equation}\label{sec4.subsec1.thm1.eq2}
P(V_{0}(t+1)=y_{0_{t+1}}\vert B(t))=\binom{x_{t}}{y_{0_{t+1}}}\left(P^{i}_V(t)\right)^{y_{0_{t+1}}}\left(1-P^{i}_V(t)\right)^{{x_{t}}-y_{0_{t+1}}},
\end{equation}
and
\begin{equation}\label{sec4.subsec1.thm1.eq3}
P(I_{0}(t+1)=z_{0_{t+1}}\vert B(t))=\binom{x_{t}}{z_{0_{t+1}}}\left(P^{i}_I(t)\right)^{z_{0_{t+1}}}\left(1-P^{i}_I(t)\right)^{{x_{t}}-z_{0_{t+1}}}.
\end{equation}
\end{thm}
Proof:\\
Let $not(V_{0}(t)\vee I_{0}(t) )$ be the random variable representing the number of people who remain susceptible at time $t$. It follows form basic probability rules that
\begin{eqnarray}
       &&P(B(t+1)=b_{t+1}\vert B(t))=P(V_{0}(t=1)=y_{0_{t+1}}\vert B(t)=b(t))\nonumber\\
       &&\times P(I_{0}(t+1)=z_{0_{t+1}}\vert V_{0}(t+1)=y_{0_{t+1}}, B(t)=b(t))\times\nonumber\\
        &&\times P(not(V_{0}(t+1)\vee I_{0}(t+1) )=x_{t}-y_{0_{t+1}}-z_{0_{t+1}}\vert I_{0}(t=1)=z_{0_{t+1}} ,V_{0}(t=1)=y_{0_{t+1}}, B(t)=b(t)).\nonumber\\
       \label{sec4.subsec1.thm1.proof1.eq1}
    \end{eqnarray}
        From the assumptions (a)-(d) above, it is easy to see that since $not(V_{0}(t+1)\vee I_{0}(t+1) )+V_{0}(t+1)+ I_{0}(t+1)=x_{t}$, then
        \begin{eqnarray}
    P(B(t+1)=b_{t+1}\vert B(t))&=&\binom{\underbrace{x_t}}{y_{0_{t+1}}}\binom{x_{t}-y_{0_{t+1}}}{z_{0_{t+1}}}\binom{x_{t}-y_{0_{t+1}}-z_{0_{t+1}}}{x_{t}-y_{0_{t+1}}-z_{0_{t+1}}}\left(P^{i}_V(t)\right)^{y_{0_{t+1}}}
    \times\nonumber\\
    &&\times\left(P^{i}_I(t)\right)^{z_{0_{t+1}}}
    \left(1-P^{i}_V(t)-P^{i}_I(t)\right)^{x_{t}-z_{0_{t+1}}-y_{0_{t+1}}}
        \label{sec4.subsec1.thm1.proof1.eq4}
\end{eqnarray}
    Note that (\ref{sec4.subsec1.thm1.proof1.eq4}) reduces to (\ref{sec4.subsec1.thm1.eq1}).

    The result for (\ref{sec4.subsec1.thm1.eq2})-(\ref{sec4.subsec1.thm1.eq3}) follows immediately from the assumptions (a)-(d) above.
\subsection{The SVIR Influenza Model with exponential time until vaccination} \label{sec4.subsec2}
In this subsection we consider a less aggressive influenza scenario, where  the state of every susceptible person in the next time step is either just infected $I_{0}$, just vaccinated $V_{0}$, or the person remains susceptible $S$. Due to limited space, we shall present only the assumptions of this special SVIR model, so as to enlighten about more realistic special cases of the SVIR model in Theorem~\ref{ch2.sec3.thm1}. The analysis of this special case will appear in the second part of this study.

 We let $P^{i}_{I}(t)$ be the probability that in the next time step $t$ the $i^{th}$ susceptible person becomes infected, and $P^{i}_{V}(t)$ be the probability that in the next time step $t$ the $i^{th}$ susceptible person becomes vaccinated, and $P^{i}_{S}(t)$ be the probability that in the next time step $t$ the $i^{th}$ susceptible person remains susceptible, where $P^{i}_{V}(t)+P^{i}_{I}(t)+P^{i}_{S}(t)=1$, we shall use the multinomial distribution to characterize the transition probabilities for the SVIR stochastic process $\{B(t), t=0,1,2,\dots\}$ defined in Theorem~\ref{ch2.sec3.thm1}. The following are slight modifications of the assumptions (a)-(d) made for the special  SVIR model in Subsection~\ref{sec4.subsec1}.

Define $A^{i}_{t}$ be an indicator random variable for the event that the $i^{th}$ susceptible person becomes infected at time $t$. Also let  $B^{i}_{t}$ be a discrete random variable denoting the number of infectious people that the $i^{th}$ susceptible person meets, or interacts with at time $t$.

(f) It is assumed for the influenza epidemic that the population is also well sensitized and educated about the disease, and there are available vaccines for those seeking vaccination, as well as some people in the population practice natural preventive techniques against the disease. Thus,  because of availability of vaccination and natural control measures, every susceptible person in the population in the next time step $t$, that is in $(t-1,t]$,  is either infected, vaccinated or remains susceptible to the disease.

(g) Similarly to (b) in Section~\ref{sec4.subsec1},  at time $t$, the probability that the $i^{th}$ susceptible individual from the $(t-1)^{th}$ generation $S(t-1)=x_{t-1}$ becomes infectious after interaction with $j$ infectious individuals is denoted $p_j^i(t) \equiv p_j^i$, and is given by
\begin{equation}\label{sec4.subsec2.eq1}
   P(A^{i}_{t}=1\vert B^{i}_{t}=j)\equiv p_j^i=1-(1-p)^j, i=1, 2, 3,\dots, x_{t-1}, \text{where} \quad S(t-1)=x_{t-1}.
\end{equation}

(h) Using ideas from \cite{novelframework}, we let $N_{i}(t)=N$ represent the number of people the $i^{th}$ susceptible individual interacts with at time $t$, where $N=0,1,2,\ldots$.
We assume that the random process $\{N_i(t), t=0, 1, 2, \dots\}$ is a poisson process with rate $\lambda$ per unit time.

(i) Furthermore, it is assumed that there is homogenous mixing in the population, and as a result, the probability that the $i^{th}$ susceptible person $S(t-1)=x_{t-1}$ at the beginning of the $t^{th}$ time interval $(t-1,t]$  meets an infectious person $I(t-1)$ also from the $(t-1)^{th}$ generation present at time $t$ is given as $\alpha^{i}_{t}=\frac{I(t-1)}{n-1}$.  Thus, the probability that the $i^{th}$ susceptible individual meets exactly $j$ infectious individuals given that (g) holds is defined by
\begin{equation}\label{sec4.subsec2.eq2}
P(B^{i}_{t}=j\vert N_i(t)=N)\equiv P_{j}^{i}(N) = \binom{N}{j}\left(\frac{I(t-1)}{n-1}\right)^{j}\left(1-\frac{I(t-1)}{n-1}\right)^{N-j},
\end{equation}
Using (e)-(h), we find $P^{i}_{I}(t)$ in the following lemma.
\begin{lemma}\label{sec4.subsec2.lemma1}
Suppose the assumptions in (f)-(i) hold, then it follows that the probability that the $i^{th}$ susceptible individual meets infectious people at time $t$, and becomes infected is
given by
\begin{equation}\label{sec4.subsec2.lemma1.eq1}
P^{i}_{I}(t)=1-e^{-p\alpha^{i}_{t}\lambda}
\end{equation}
\end{lemma}
Proof:\\
The probability that the $i^{th}$ susceptible individual meets infectious people at time $t$, and becomes infected is given by applying multiplication rule in the following
\begin{eqnarray}
P^{i}_{I}(t)&=&P(A^{i}_{t}=1)=\sum^{\infty}_{N=0}\sum^{N}_{j=0}P(A^{i}_{t}=1,B^{i}_{t}=j,N_i(t)=N )\nonumber\\
&=&\sum^{\infty}_{N=0}\sum^{N}_{j=0}\left(1-(1-p)^j\right)\binom{N}{j}\left(\frac{I(t-1)}{n-1}\right)^{j}\left(1-\frac{I(t-1)}{n-1}\right)^{N-j}\times \nonumber\\
&&\times \frac{(\lambda(t-(t-1)))^{N}e^{-\lambda(t-(t-1))}}{N!}.\label{sec4.subsec2.lemma1.proof.eq1}
\end{eqnarray}
Applying the binomial theorem and  Taylor series expansion of the exponential function and further algebraic modifications and simplifications, the right-hand-side of (\ref{sec4.subsec2.lemma1.proof.eq1}) reduces to  (\ref{sec4.subsec2.lemma1.eq1}), where
\begin{equation}\label{sec4.subsec2.lemma1.proof.eq2}
\alpha^{i}_{t}=\frac{I(t-1)}{n-1}.
\end{equation}

The following assumptions are made to derive $P^{i}_{V}(t)$ - the probability that in the next time step $t$ the $i^{th}$ susceptible person becomes vaccinated.
Note that in some influenza scenarios, people get vaccinated  independently, and at a constant rate $\mu_{v}$ per unit time. In other words, the number of people getting vaccinated  over time follows a poisson process $\{M(t),t=0,1,2,\ldots\}$ with rate $\mu_{e}$. Thus, the time $T_{v}>0$ until the $i^{th}$ susceptible person gets vaccinated follows the exponential distribution with mean $\frac{1}{\mu_{v}}$. Very easily, the probability that the $i^{th}$ susceptible person gets vaccinated in any interval $(t-1,t]$, can be calculated using the poisson process as follows;
\begin{eqnarray}\label{D1}
P^{i}_{V}(t)=1-P(T_{v}> t|T_{v}>t-1)=1-P(M(t)-M(t-1)=0)=1-e^{-\mu_{v}}.
\end{eqnarray}
We now use the trinomial distribution to characterize the transition probabilities of the SVIR stochastic process in the following:
\begin{thm}\label{sec4.subsec2.thm1}
Let the conditions of Theorem~\ref{ch2.sec3.thm1} be satisfied, and let the probability that the $i^{th}\forall i=1,2,\ldots,x_{t}$ susceptible gets vaccinated at time $t$, denoted $P^{i}_V(t)$ be as defined in (\ref{D1}), the probability that the $i^{th}\forall i=1,2,\ldots,x_{t}$ susceptible gets infected at time $t$, denoted $P^{i}_I(t)$ be  as defined in (\ref{sec4.subsec2.lemma1.eq1}), where $P^{i}_{V}(t)+P^{i}_{I}(t)+P^{i}_{S}(t)=1$, and $P^{i}_{S}(t)$ is the probability that in the next time step $t$ the $i^{th}$ susceptible person remains susceptible. It follows that for $t\geq max\{T_1, T_2\}$, the transition probabilities for the stochastic process $\{B(t), t=1,2,3,\dots\}$, are given as follows:
%
\begin{eqnarray}
P(B(t+1)=b(t+1)\vert B(t)=b(t))&=&P(V_{0}(t+1) = y_{0_{t+1}}, I_0(t+1) = z_{0_{t+1}} \vert B(t))\nonumber \\
&=&\frac{x_{t}!}{y_{0_{t+1}}!z_{0_{t+1}}!(x_{t} - y_{0_{t+1}}-z_{0_{t+1}})!}\times \nonumber \\
&&(P^{i}_{V}(t))^{y_{0_{t+1}}}(P^{i}_{I}(t))^{z_{0_{t+1}}}(1-P^{i}_{V}(t)-P^{i}_{I}(t))^{x_{t}-y_{0_{t+1}}-z_{0_{t+1}}},\nonumber \\
\label{4.3}
\end{eqnarray}
where $P^{i}_{V}(t)$ and $P^{i}_{I}(t)$ are defined above.  Moreover, it follows from (\ref{4.3}) that the conditional marginal distributions of $V_0$ and $I_0$ are binomials and given as follows:
\begin{eqnarray}\label{4.3a}
    P(V_0(t+1) = y_{0_{t+1}}\vert B(t)) = \binom{x_t}{y_{0_{t+1}}}(P^{i}_{V}(t))^{y_{0_{t+1}}}(1-P^{i}_{V}(t))^{x_t-y_{0_{t+1}}};\nonumber \\
    y_{0_{t+1}} = 0,1,2,\dots, x_t,
\end{eqnarray}
\begin{eqnarray}\label{4.3b}
    P(I_0(t+1)=z_{0_{t+1}}\vert B(t)=b(t)) = \binom{x_t}{z_{0_{t+1}}}(P^{i}_{I}(t))^{z_{0_{t+1}}}(1-P^{i}_{I}(t))^{x_t+z_{0_{t+1}}}; \nonumber \\
    z_{0_{t+1}}=0,1,2,\dots, x_t.
\end{eqnarray}
where $y_{0_{t+1}}=0,1,2,\ldots,\underbrace{x_{t}}$ and $z_{0_{t+1}}=0,1,2,\ldots,{x_{t}}-y_{0_{t+1}}$.
\end{thm}
Proof:\\
The result follows easily by applying the trinomial distribution to the conditional distribution $P(V_{0}(t+1) = y_{0_{t+1}}, I_0(t+1) = z_{0_{t+1}} \vert B(t))$.

The analysis of the special SVIR model in Theorem~\ref{sec4.subsec2.thm1} appears in the part 2 of this study.
\section{Parameter estimation}\label{sec4.sec5}
In this section, we find estimators for the true parameters of the SVIR Markov chain model using observed data for the state of the process over time.  In Subsection~\ref{sec4.subsec1}, observe that both  $P^{i}_V(t)$  and $P^{i}_I(t)$ depend on the probability of getting infection from one interaction with an infectious person, $p$. Also, $P^{i}_V(t)$ depends on $\phi$- the probability that a susceptible person who has eluded infection develops desire and become vaccinated at time $t$.  Therefore, utilizing ideas from \cite{novelframework}, we find a maximum likelihood estimator (MLE's) for the probability of getting infection from one interaction with an infectious person, $p$, and also for $\phi$, for the chain $\{B(t), t=0,1,2,\dots\}$ in the case defined in Subsection~\ref{sec4.subsec1}, that is, when the transition probabilities are defined in (\ref{sec4.subsec1.thm1.eq1}).

  Furthermore, we consider the random process $\{B(t), t=0,1,2,\dots\}$, whenever $t\geq \max\{T_1, T_2\}$ without loss of generality. Also, note that the parameter $\Theta = (p,\phi)$ represent fixed measurements in the population at each time $t$, that is, $p$ and $\phi$ represent  fixed measurements for events occurring in the population during the $t^{th}$ time interval $(t-1, t]$, where the population at any time $t=0, 1, 2, \ldots$ is defined by the random vector
\begin{equation}\label{sec4.sec5.eq1}
B(t)=(S(t), V_0(t), V_1(t), \dots, V_{T_1-1}(t), I_0(t), I_1(t), \dots, I_{T_2-1}(t)),
\end{equation}
whenever $t\geq \max\{T_1, T_2\}$.

Note that due to the size of this paper, investigations and  MLE's for the parameters $\Theta = (p, \lambda, r, k)$ for the SVIR chain-binomial model in Subsection~\ref{sec4.subsec2} will appear in the Part-2 of this work.

Let $\hat{b}(t)$ be the observed value of the random vector $B(t)$ at time $t\geq \max\{T_1, T_2\}$, $t=0,1,2,\dots$, defined in (\ref{ch2.sec3.eq6a}).  That is,
\begin{equation}\label{C0}
    \hat{b}(t)=(\hat{x}_t=\underbrace{\hat{x}_t}+\overbrace{\hat{x}_t}, \hat{y_{0_t}}, \hat{y_{1_t}}, \hat{y_{2_t}}, \dots, \hat{y_{{T_1-1}_t}}, \hat{z_{0_t}}, \hat{z_{1_t}}, \hat{z_{2_t}}, \dots, \hat{z_{{t_2-1}_t}} ),
\end{equation}
where $\hat{x}_t, \hat{y_{0_t}}, \hat{y_{1_t}}, \hat{y_{2_t}}, \dots, \hat{y_{{T_1-1}_t}}, \hat{z_{0_t}}, \hat{z_{1_t}}, \hat{z_{2_t}}, \dots, \hat{z_{{t_2-1}_t}} \in \mathbb{Z}_+$ are non-negative observed constant values for each component of $B(t)$, at time $t=0,1,2,3,\dots$.

The population $B(t)$ is observed over the time units, $t=0,1,2,\dots, T$, where the initial state $B(0) = \hat{b}(0)$ is assumed to be known.  That is, $B(0)$ is deterministic, and the observed data consists of the measurements
\begin{equation}\label{D0}
    \hat{b}(0), \hat{b}(1), \hat{b}(2), \dots, \hat{b}(T).
\end{equation}
We define the collection of random variables ${B}(0), {B}(1), {B}(2), \dots, {B}(T)$ representing the population over times $t=0,1,2,\dots, T$ as follows:
\begin{equation}\label{E0}
    D_T=\{ B(0), B(1), B(2), \dots, B(T)\}
\end{equation}
and from (\ref{C0}), the observed values of $D_T$ are given as
\begin{equation}\label{F0}
    \hat{D}_T=\{\hat{b}(0), \hat{b}(1), \hat{b}(2), \dots, \hat{b}(T)\}.
\end{equation}
We use the observed sample path $\hat{D}_T$ of the process $\{B(t): t=0,1,2,\dots\}$ to find maximum likelihood estimators for the parameters $\Theta=(p,\phi)$.  The generation of the sample path $\hat{D}_T$ in (\ref{F0}) from the population $B(t)$ over times $t=0,1,2,\dots, T$ is exhibited in Figure~\ref{F2}.

\begin{figure}[h]
\begin{center}

\caption{Shows the transition of the process $\{B(t), t=0, 1, 2, \dots\}$ over time $t=0, 1, 2, \dots, T$, and observed data $\hat{D}_T = \{\hat{b}(0), \hat{b}(1), \hat{b}(2), \dots, \hat{b}(T)\}$.  The parameters $\Theta = (p, \phi)$ are constant in the population at all times $t=0, 1, 2, \dots, T$.}\label{F2}

\end{center}
\end{figure}

\begin{rem}
It must be noted that inferences for the parameters $\Theta=(p,\phi)$ such as confidence intervals and tests of significance, and consistency of estimators for the parameters are beyond the scope of this work, and will appear elsewhere.
\end{rem}

We assume that we have data for influenza over time units $t=0,1,2,\dots, T$ denoted $\hat{D}_T$, where $\hat{D}_T$ is defined in (\ref{F0}), and $\hat{D}_T$ is one realization of the human population over time denoted $D_T$, defined in (\ref{E0}).
 From  (\ref{C0}), (\ref{E0}), and (\ref{F0}), the likelihood function of $\Theta=(p,\phi)$ is defined as follows:

\begin{eqnarray}\label{5.3}
    L(\Theta \vert \hat{D}_T)&=&L(p\vert \hat{D}_T) = P(D_T=\hat{D}_T\vert p,\phi) \nonumber \\
    &&=P(B(T)=\hat{b}(T), B(T-1)=\hat{b}(T-1), \dots, B(0)=\hat{b}(0)\vert p,\phi).\nonumber \\
    \end{eqnarray}
From (\ref{5.3}), applying the multiplication rule, it is easy to see that
\begin{eqnarray}\label{H0}
    L(p\vert D)&&=P(B(T)=\hat{b}(T)\vert B(T-1)=\hat{b}(T-1), \dots, B(0)=\hat{b}(0);  p,\phi) \times \nonumber \\
    &&\times P(B(T-1)=\hat{b}(T-1)\vert B(T-2)=\hat{b}(T-2), \dots, B(0)=\hat{b}(0);  p,\phi)\times \nonumber \\
    &\vdots& \nonumber \\
    &&\times P(B(1)=\hat{b}(1)\vert B(0)=\hat{b}(0);p_v, p)\times P(B(0)=\hat{b}(0); p,\phi).
\end{eqnarray}
But, since $\{B(t), t=0,1,2,3,\dots\}$ is a Markov Chain, and since it is assumed $B(0)$ is known, it is easy to see that (\ref{H0}) reduces to
\begin{equation}\label{J0}
    L(p\vert\hat{D}_T) = \prod_{k=1}^{T}P(B(k)=\hat{b}(k)\vert B(k-1)=\hat{b}(k-1);p,\phi).\\
\end{equation}
It follows from (\ref{J0}) and (\ref{ch2.sec3.thm1.eq2}) that for $S(k)=x_{k}\in (0,x_{k-1}),\forall k=1,2,\ldots, T$,
\begin{equation}\label{K0}
    L( p\vert \hat{D}_T)=\prod_{k=1}^{T}P(V_0(k)=y_{0_k}, I_0(k)=z_{0_k}\vert B(k-1)=\hat{b}(k-1);  p,\phi). \\
\end{equation}
Observe that substituting (\ref{sec4.subsec1.thm1.eq1}) into (\ref{K0}) leads to the likelihood function with respect to the parameters $p,\phi$.
We should note that applying the maximization technique to find the MLE $\hat{p}$ for $p$ using the likelihood function $L$ defined in (\ref{K0}) leads to an intractable equation for the derivative of the log-likelihood of $L$.  Thus, we apply the expectation maximization (EM) algorithm to find an appropriate MLE for $p$- the probability of passing the infection from one independent infectious contact.
\subsection{The EM Algorithm and Jensen's Inequality}\label{sec4.sec6}
In this section, we consider the EM algorithm, where missing information is incorporated into the incomplete likelihood function, at random, and  Jensen's inequality is used to find a lower-bound for the complete log-likelihood function.

Recall, the Expectation Maximization (EM) algorithm is an iterative algorithm used to find the MLE of a parameter $\Theta$ of a given distribution\cite{EM, gentleEM}.  There are two cases where the algorithm is most useful: (1) when the data available for the maximum likelihood estimation technique has missing components, and (2) when maximizing the likelihood function leads to an intractable equation,  but adding missing data can simplify the process.  It is for the second case in our problem that we utilize this EM algorithm.

Suppose we have observed data $Y,$ and likelihood function $L(\Theta \vert Y) = P(Y \vert \Theta)$, and suppose the vector $Z$ is missing data or a missing component, so that $X=(Y,Z)$ is the complete data.  The complete log-likelihood function $\log L(\Theta \vert X) = \log P(Y,Z \vert \Theta)$ can be maximized to find the MLE of $\Theta$ in two basic EM algorithm steps, namely - the expectation (E) step, and the maximization (M)-step.

The E-step consists of finding the expected value of the complete log-likelihood function $\log P(Y,Z \vert \Theta)$ with respect to the conditional mass of $Z$ given $Y$ and $\Theta$.  That is, we find
\begin{eqnarray}
E_{Z\vert Y;\Theta}[\log L(\Theta \vert X)]&=&E_{Z \vert Y;\Theta}[\log P(Y,Z\vert \Theta)]\nonumber \\
&=&\sum_{Z}\log (P(Y,Z\vert \Theta)P(Z\vert Y; \Theta)).\label{sec4.sec6.eq1}
\end{eqnarray}

The M-step consists of maximizing $E_{Z\vert Y;\Theta}[\log L(\Theta \vert X)]$ to find an estimate $\hat{\Theta}$ for $\Theta$.  This process is summarized in the following steps:
\begin{itemize}
    \item[i.] Let $m=0$ and $\hat{\Theta}^m$ be an initial guess for $\Theta$. \\
    \item[ii.] Given the observed data $Y$, and assuming that the guess $\hat{\Theta}^m$ is correct, calculate the conditional probability distribution $P(Z\vert Y, \hat{\Theta}^m)$ for the missing data $Z$.\\
    \item[iii.] Find the conditional expected log-likelihood referred to as $Q$, that is,
    \begin{eqnarray}
        Q(\Theta \vert \hat{\Theta}^m) &=& \sum_{Z} \log (P(Y,Z \vert \Theta)P(Z \vert Y,\hat{\Theta}^m))\nonumber \\
        &=& E_{Z\vert Y,\hat{\Theta}^m}[\log (P(X\vert \Theta))],
    \end{eqnarray}\\
    where $X=(Y,Z)$.
    \item[iv.] Find the $\Theta$ that maximizes $Q(\Theta \vert \hat{\Theta}^m)$.  The result will be the new $\hat{\Theta}^{m+1}$.  That is, \\
    \begin{equation}
        \hat{\Theta}^{m+1}=\text{argmax}_{\Theta}Q(\Theta\vert \hat{\Theta}^m).
    \end{equation}
    \item[v.] Update $\hat{\Theta}^m$ and repeat step (iv) until $\Theta$ stops noticeably changing.
\end{itemize}
The  E-step can be obtained by applying Jensen's inequality.  We recall Jensen's inequality \cite{burgers} in the following:
 \begin{lemma}\label{sec4.sec6.lemma1}
 Suppose $f$ is a convex function, and $X$ is a random variable, then
\begin{equation}\label{sec4.sec6.lemma1.eq1}
    E[f(X)] \geq f(E[X]).
\end{equation}
Conversely, if you have a concave function (e.g. a logarithmic function), then
\begin{equation}\label{sec4.sec6.lemma1.eq2}
    E[f(x)]\leq f(E[X]).
\end{equation}
\end{lemma}
%

From (\ref{sec4.sec6.eq1}), let $Y=\hat{D}_T$ represent the observed data defined in (\ref{F0}).  The following random missing information $Z$ are incorporated to make the log-likelihood function $log(L)$ more tractable, where $L$ is given in (\ref{K0}).
\begin{description}
    \item[(i.)] the collection $\vec{e}_T^i = \{e_0^i, e_1^i, e_2^i, \ldots,e_t^i,\ldots, e_T^i\}$, where for each $t \in \{0, 1, 2, \dots, T\}$ and $i \in \{ 1, 2, 3, \dots, x_{t-1}=S(t-1)\}$, $e_t^i$ is a discrete random variable representing the number of infectious individuals that the $i^{th}$ susceptible person of the $(t-1)^{th}$ generation meets at time $t$ (i.e. in the interval $(t-1, t])$.  Since from (\ref{B8}), a susceptible person only meets the fixed number of people, $N$, at any time $t$, therefore
    \begin{equation}\label{K1}
        e_t^i = j, j=1,2,3,\dots, N.
    \end{equation}
    \item[(ii.)] the collection $\vec{d}_{T,j}^i = \{d_{0,j}^i, d_{1,j}^i, d_{2,j}^i, \ldots,d_{t,j}^i, \ldots, d_{T-1, j}^i, d_{T,j}^i\}$, where for each $t \in \{0,1,2,\dots, T\}$, $i \in \{1,2,3,\dots, x_{t-1}=S(t-1)\}$, and $j \in \{1,2,3,\dots, N\}$, $d_{t,j}^i$ is a categorical random variable indicating the event that the $l^{th}$ infectious individual passes the infection with probability $p$, where $l = 1,2,3,\dots, j$, given that the $i^{th}$ susceptible person meets $j$ infectious people among the fixed number of people $N$, at time $t$.
        Thus,
    \begin{equation}\label{K11}
        d_{t,j}^i = l, l = 1,2,3,\dots, j.
    \end{equation}
\end{description}

We consider a step-by-step approach to add the missing data $\vec{e}_T^i$ and $\vec{d}_{T,j}^i$ into the incomplete likelihood function $L$.
\begin{lemma}\label{sec4.sec6.lemma1}
Let Theorem~\ref{sec4.subsec1.thm1} hold. Given the missing information $\vec{e}_T^i$ and $\vec{d}_{T,j}^i$ defined in (\ref{K1}) and (\ref{K11}), then it follows from (\ref{sec4.subsec1.thm1.eq1}) and (\ref{K0}),  that the log-likelihood function $L(p,\phi \vert \hat{D}_T)$ satisfies the following inequality:
\begin{eqnarray}
    \log L( p\vert \hat{D}(t)) &\geq& \sum_{k=1}^{T}\sum_{j=1}^{N}\sum_{l=1}^{j}\left[P(e_k^i=j \vert \hat{B}(k-1)=\hat{b}(k-1); p)\times \right. \nonumber \\
    &&\left. \times P(d_{k,j}^i=l \vert e_k^i = j, \hat{B}(k-1)=\hat{b}(k-1);p) \times \right. \nonumber \\
    &&\left. \times \log (P(V_0(k)=\hat{y}_{0_k}, e_k^i=j, d_{k, j}^i=l\vert \hat{B}(k-1)=\hat{b}(k-1); p))\right] \nonumber \\
    &&+\sum_{k=1}^{T}\sum_{j=1}^{N}\sum_{l=1}^{j}\left[P(e_k^i=j \vert V_0(k)=\hat{y}_{0_k}, \hat{B}(k-1)=\hat{b}(k-1); p)\times \right. \nonumber \\
    &&\left. \times P(d_{k,j}^i=l \vert e_k^i = j,V_0(k)=\hat{y}_{0_k}, \hat{B}(k-1)=\hat{b}(k-1);p) \times \right. \nonumber \\
    &&\left. \times \log (P(I_0(k)=\hat{y}_{0_k}, e_k^i=j, d_{k, j}^i=l\vert V_0(k)=\hat{y}_{0_k}, \hat{B}(k-1)=\hat{b}(k-1); p))\right] \nonumber \\
    &&-\lambda_{1}-\lambda_{2}\nonumber\\
    &&+\sum_{k=1}^{T}\sum_{j=1}^{N}\sum_{l=1}^{j}\left[P(e_k^i=j \vert I_0(k)=\hat{z}_{0_k},V_0(k)=\hat{y}_{0_k}, \hat{B}(k-1)=\hat{b}(k-1); p,\phi)\times \right. \nonumber \\
    &&\left. \times P(d_{k,j}^i=l \vert e_k^i = j,I_0(k)=\hat{z}_{0_k},V_0(k)=\hat{y}_{0_k}, \hat{B}(k-1)=\hat{b}(k-1);p,\phi) \times \right. \nonumber \\
    &&\left. \times \log (P(not(V_0(k)\vee I_0(k))=\hat{u}_{0_k}, e_k^i=j, d_{k, j}^i=l\vert I_0(k)=\hat{z}_{0_k},V_0(k)=\hat{y}_{0_k}, \hat{B}(k-1)=\hat{b}(k-1); p,\phi))\right] \nonumber \\
    &&-\lambda_{1}-\lambda_{2}\label{sec4.sec6.lemma1.eq1}
\end{eqnarray}
where $\hat{u}_{0_k}=\hat{x}_{k-1}-\hat{y}_{0_k}-\hat{z}_{0_k},\forall k=1,2,\ldots, T$, and  $\lambda_{1}$,  and also $\lambda_{2}$ are  probability terms that depends on $\vec{e}_T^i$ and $\vec{d}_{T,j}^i$.
\end{lemma}
Proof:\\
From (\ref{K0}), denote the log-likelihood $l(p,\phi\vert \hat{D}(t))\equiv\log{L( p,\phi\vert \hat{D}(t))}$. It follows from (\ref{K0})  that adding the missing data $\vec{e}_T^i$, we obtain
\begin{eqnarray}
l( p,\phi\vert \hat{D}(t))&=& \log{\prod_{k=1}^{T}P(V_0(k)=y_{0_k}, I_0(k)=z_{0_k}\vert B(k-1)=\hat{b}(k-1);  p,\phi)} \nonumber \\
&=& \sum_{k=1}^{T}\log\left[\sum_{j=1}^{N}P(V_0(k)=y_{0_k}, e_k^i=j \vert \hat{B}(k-1)=\hat{b}(k-1); p,\phi)\right]\nonumber \\
&&+ \sum_{k=1}^{T}\log\left[\sum_{j=1}^{N}P(I_0(k)=z_{0_k}, e_k^i=j \vert V_0(k)=y_{0_k,} \hat{B}(k-1)=\hat{b}(k-1); p,\phi)\right]\nonumber \\
&&+ \sum_{k=1}^{T}\log\left[\sum_{j=1}^{N}P(not(V_0(k)\vee I_0(k))=u_{0_k}, e_k^i=j \vert I_0(k)=z_{0_k,}, V_0(k)=y_{0_k,} \right.\nonumber\\
&&\left. \hat{B}(k-1)=\hat{b}(k-1); p,\phi)\right].\nonumber \\
\label{sec4.sec6.lemma1.proof.eq1}
\end{eqnarray}
The equation (\ref{sec4.sec6.lemma1.proof.eq1}) can be expressed further as follows
\begin{eqnarray}
l( p,\phi\vert \hat{D}(t))
&=& \sum_{k=1}^{T}\log\left[\sum_{j=1}^{N}\frac{P(V_0(k)=y_{0_k}, e_k^i=j \vert \hat{B}(k-1)=\hat{b}(k-1))}{P( e_k^i=j \vert \hat{B}(k-1)=\hat{b}(k-1); p)}; p,\phi)\times \right.\nonumber\\
&&\times \left.P( e_k^i=j \vert \hat{B}(k-1)=\hat{b}(k-1);p,\phi)\right]\nonumber \\
&&+ \sum_{k=1}^{T}\log\left[\sum_{j=1}^{N}\frac{P(I_0(k)=y_{0_k}, e_k^i=j \vert V_0(k)=y_{0_k,} \hat{B}(k-1)=\hat{b}(k-1); p,\phi)}{P( e_k^i=j \vert V_0(k)=y_{0_k,} \hat{B}(k-1)=\hat{b}(k-1) p,\phi)}\times \right.\nonumber\\
&&\times \left. P( e_k^i=j \vert V_0(k)=y_{0_k,} \hat{B}(k-1)=\hat{b}(k-1); p,\phi)\right].\nonumber \\
&&+ \sum_{k=1}^{T}\log\left[\sum_{j=1}^{N}\frac{P(not(V_0(k)\vee I_0(k))=u_{0_k}, e_k^i=j \vert I_0(k)=z_{0_k,},V_0(k)=y_{0_k}, \hat{B}(k-1)=\hat{b}(k-1);p,\phi)}{P( e_k^i=j \vert V_0(k)=y_{0_k,} \hat{B}(k-1)=\hat{b}(k-1);p,\phi)}\times \right.\nonumber\\
&&\times \left. P( e_k^i=j \vert I_0(k)=z_{0_k,},V_0(k)=y_{0_k,} \hat{B}(k-1)=\hat{b}(k-1); p,\phi)\right].\nonumber \\
\label{sec4.sec6.lemma1.proof.eq2}
\end{eqnarray}
Applying Jensen's inequity to (\ref{sec4.sec6.lemma1.proof.eq2}), leads to the following
 \begin{eqnarray}
l( p,\phi\vert \hat{D}(t))
&\geq & \sum_{k=1}^{T}\sum_{j=1}^{N} \log{\left\{P(V_0(k)=y_{0_k}, e_k^i=j \vert \hat{B}(k-1)=\hat{b}(k-1);p,\phi)\right\}}P( e_k^i=j \vert \hat{B}(k-1)=\hat{b}(k-1); p,\phi)\nonumber\\
&&+\sum_{k=1}^{T}\sum_{j=1}^{N}\log{\left\{P(I_0(k)=z_{0_k}, e_k^i=j \vert V_0(k)=y_{0_k,} \hat{B}(k-1)=\hat{b}(k-1); p,\phi)\right\}}\times\nonumber\\
&&\times P( e_k^i=j \vert V_0(k)=y_{0_k,} \hat{B}(k-1)=\hat{b}(k-1); p,\phi)\nonumber\\
&&+\sum_{k=1}^{T}\sum_{j=1}^{N}\log{\left\{P(not(V_0(k)\vee I_0(k))=u_{0_k}, e_k^i=j \vert I_0(k)=z_{0_k,},V_0(k)=y_{0_k}, \hat{B}(k-1)=\hat{b}(k-1);p,\phi)\right\}}\times\nonumber\\
&&\times P( e_k^i=j \vert I_0(k)=z_{0_k}, V_0(k)=y_{0_k}, \hat{B}(k-1)=\hat{b}(k-1); p,\phi)\nonumber\\
&&-\lambda_{1},
\label{sec4.sec6.lemma1.proof.eq3}
\end{eqnarray}
where (\ref{sec4.sec6.lemma1.proof.eq3}) $\lambda_{1}$ is given as follows
\begin{eqnarray}
\lambda_{1}
&= & \sum_{k=1}^{T}\sum_{j=1}^{N} \log{\left\{P( e_k^i=j \vert \hat{B}(k-1)=\hat{b}(k-1); p,\phi)\right\}}P( e_k^i=j \vert \hat{B}(k-1)=\hat{b}(k-1); p,\phi)\nonumber\\
&&+\sum_{k=1}^{T}\sum_{j=1}^{N}\log{\left\{P( e_k^i=j \vert V_0(k)=y_{0_k,} \hat{B}(k-1)=\hat{b}(k-1); p,\phi)\right\}}\times\nonumber\\
&&\times P( e_k^i=j \vert V_0(k)=y_{0_k,} \hat{B}(k-1)=\hat{b}(k-1); p,\phi)
\nonumber\\
&&+\sum_{k=1}^{T}\sum_{j=1}^{N}\log{\left\{P( e_k^i=j \vert I_0(k)=z_{0_k}, V_0(k)=y_{0_k}, \hat{B}(k-1)=\hat{b}(k-1); p,\phi)\right\}}\times\nonumber\\
&&\times P( e_k^i=j \vert I_0(k)=z_{0_k}, V_0(k)=y_{0_k}, \hat{B}(k-1)=\hat{b}(k-1); p,\phi)\nonumber\\
\label{sec4.sec6.lemma1.proof.eq4}
\end{eqnarray}
 We add the missing data $d_{T,j}^i$ in (\ref{K11}) into the partially complete log likelihood function $\log{P(V_0(k)=y_{0_k}, e_k^i=j \vert \hat{B}(k-1)=\hat{b}(k-1);p,\phi)} $, \\ $\log{P(I_0(k)=y_{0_k}, e_k^i=j \vert V_0(k)=y_{0_k,} \hat{B}(k-1)=\hat{b}(k-1); p,\phi)}$ and\\ $\log{\left\{P(not(V_0(k)\vee I_0(k))=u_{0_k}, e_k^i=j \vert I_0(k)=z_{0_k,},V_0(k)=y_{0_k}, \hat{B}(k-1)=\hat{b}(k-1);p,\phi)\right\}}$,
 $\forall k \in \{1,2,3,\dots,T\}$; $j\in \{1,2,3,\dots, N\}$, and apply the same technique in (\ref{sec4.sec6.lemma1.proof.eq1})-(\ref{sec4.sec6.lemma1.proof.eq4}), as follows. 
From (\ref{sec4.sec6.lemma1.proof.eq3})
\begin{eqnarray}
l( p,\phi\vert \hat{D}(t))
&\geq & \sum_{k=1}^{T}\sum_{j=1}^{N} \log{\left\{\sum^{j}_{l=1}P(V_0(k)=y_{0_k}, e_k^i=j, d_{k,j}^i=l \vert \hat{B}(k-1)=\hat{b}(k-1))\right\}}\times\nonumber\\
&&\times P( e_k^i=j \vert \hat{B}(k-1)=\hat{b}(k-1); p,\phi)\nonumber\\
&&+\sum_{k=1}^{T}\sum_{j=1}^{N}\log{\left\{\sum^{j}_{l=1}P(I_0(k)=y_{0_k}, e_k^i=j,d_{k,j}^i=l \vert V_0(k)=y_{0_k,} \hat{B}(k-1)=\hat{b}(k-1); p,\phi)\right\}}\times\nonumber\\
&&\times P( e_k^i=j \vert V_0(k)=y_{0_k,} \hat{B}(k-1)=\hat{b}(k-1); p,\phi)\nonumber\\
&&+\sum_{k=1}^{T}\sum_{j=1}^{N}\log{\left\{\sum^{j}_{l=1}P(not(V_0(k)\vee I_0(k))=u_{0_k}, e_k^i=j,d_{k,j}^i=l \vert I_0(k)=z_{0_k,}, V_0(k)=y_{0_k}, \hat{B}(k-1)=\hat{b}(k-1); p,\phi)\right\}}\times\nonumber\\
&&\times P( e_k^i=j \vert I_0(k)=z_{0_k}, V_0(k)=y_{0_k,} \hat{B}(k-1)=\hat{b}(k-1); p,\phi)
\nonumber\\
&&-\lambda_{1},
\label{sec4.sec6.lemma1.proof.eq5}
\end{eqnarray}
Applying Jensen's inequality on (\ref{sec4.sec6.lemma1.proof.eq5}) we obtain the following
\begin{eqnarray}
&&l( p,\phi\vert \hat{D}(t))
\geq  \sum_{k=1}^{T}\sum_{j=1}^{N} \sum^{j}_{l=1}\log{\left\{P(V_0(k)=y_{0_k}, e_k^i=j, d_{k,j}^i=l \vert \hat{B}(k-1)=\hat{b}(k-1);p,\phi)\right\}}\times\nonumber\\
&&\times P( d_{k,j}^i=l\vert e_k^i=j, \hat{B}(k-1)=\hat{b}(k-1); p,\phi)P( e_k^i=j \vert \hat{B}(k-1)=\hat{b}(k-1); p,\phi)\nonumber\\
&&+\sum_{k=1}^{T}\sum_{j=1}^{N}\sum^{j}_{l=1}\log{\left\{P(I_0(k)=y_{0_k}, e_k^i=j,d_{k,j}^i=l \vert V_0(k)=y_{0_k}, \hat{B}(k-1)=\hat{b}(k-1);p,\phi)\right\}}\times\nonumber\\
&&\times P(d_{k,j}^i=l \vert e_k^i=j, V_0(k)=y_{0_k}, \hat{B}(k-1)=\hat{b}(k-1); p,\phi) P( e_k^i=j \vert V_0(k)=y_{0_k}, \hat{B}(k-1)=\hat{b}(k-1); p,\phi)\nonumber\\
&&+\sum_{k=1}^{T}\sum_{j=1}^{N}\sum^{j}_{l=1}\log{\left\{P(not(V_0(k)\vee I_0(k))=u_{0_k}, e_k^i=j,d_{k,j}^i=l \vert I_0(k)=z_{0_k},V_0(k)=y_{0_k}, \hat{B}(k-1)=\hat{b}(k-1); p,\phi)\right\}}\times\nonumber\\
&&\times P(d_{k,j}^i=l \vert e_k^i=j, I_0(k)=z_{0_k},V_0(k)=y_{0_k}, \hat{B}(k-1)=\hat{b}(k-1); p,\phi) P( e_k^i=j \vert I_0(k)=z_{0_k},V_0(k)=y_{0_k}, \hat{B}(k-1)=\hat{b}(k-1); p,\phi)\nonumber\\
&&-\lambda_{1}-\lambda_{2},
\label{sec4.sec6.lemma1.proof.eq6}
\end{eqnarray}
where
\begin{eqnarray}
\lambda_{2}
&= & \sum_{k=1}^{T}\sum_{j=1}^{N} \log{\left\{P( d_{k,j}^i=l\vert e_k^i=j, \hat{B}(k-1)=\hat{b}(k-1); p,\phi)\right\}}\times\nonumber\\
&&\times P( d_{k,j}^i=l\vert e_k^i=j, \hat{B}(k-1)=\hat{b}(k-1); p,\phi)P( e_k^i=j \vert \hat{B}(k-1)=\hat{b}(k-1); p,\phi)\nonumber\\
&&+\sum_{k=1}^{T}\sum_{j=1}^{N}\log{\left\{P(d_{k,j}^i=l \vert e_k^i=j, V_0(k)=y_{0_k,} \hat{B}(k-1)=\hat{b}(k-1); p,\phi)\right\}}\times\nonumber\\
&&\times P(d_{k,j}^i=l \vert e_k^i=j, V_0(k)=y_{0_k,} \hat{B}(k-1)=\hat{b}(k-1); p,\phi) \times\nonumber\\
&&\times P( e_k^i=j \vert V_0(k)=y_{0_k,} \hat{B}(k-1)=\hat{b}(k-1); p,\phi)\nonumber\\
&&+\sum_{k=1}^{T}\sum_{j=1}^{N}\sum^{j}_{l=1}\log{\left\{P(d_{k,j}^i=l \vert e_k^i=j, I_0(k)=z_{0_k},V_0(k)=y_{0_k}, \hat{B}(k-1)=\hat{b}(k-1); p,\phi)\right\}}\times\nonumber\\
&&\times P(d_{k,j}^i=l \vert e_k^i=j, I_0(k)=z_{0_k},V_0(k)=y_{0_k}, \hat{B}(k-1)=\hat{b}(k-1); p,\phi) P( e_k^i=j \vert I_0(k)=z_{0_k},V_0(k)=y_{0_k}, \hat{B}(k-1)=\hat{b}(k-1); p,\phi)\nonumber\\
\label{sec4.sec6.lemma1.proof.eq7}
\end{eqnarray}

\begin{rem}\label{sec4.sec6.rem1}
We note from (\ref{sec4.sec6.lemma1}) that the E-step of the EM algorithm consists of finding the conditional expectation term
\begin{eqnarray}
    &&Q(\Theta \vert \hat{\Theta}^m ) = \sum_{k=1}^{T}\sum_{j=1}^{N}\sum_{l=1}^{j}\left[P(e_k^i=j \vert \hat{B}(k-1)=\hat{b}(k-1); p^{m},\phi^{m})\times \right. \nonumber \\
    &&\left. \times P(d_{k,j}^i=l \vert e_k^i = j, \hat{B}(k-1)=\hat{b}(k-1);p^{m},\phi^{m}) \times \right. \nonumber \\
    &&\left. \times \log (P(V_0(k)=\hat{y}_{0_k}, e_k^i=j, d_{k, j}^i=l\vert \hat{B}(k-1)=\hat{b}(k-1); p,\phi))\right] \nonumber \\
    &&+\sum_{k=1}^{T}\sum_{j=1}^{N}\sum_{l=1}^{j}\left[P(e_k^i=j \vert V_0(k)=\hat{y}_{0_k}, \hat{B}(k-1)=\hat{b}(k-1); p^{m},\phi^{m})\times \right. \nonumber \\
    &&\left. \times P(d_{k,j}^i=l \vert e_k^i = j,V_0(k)=\hat{y}_{0_k}, \hat{B}(k-1)=\hat{b}(k-1);p^{m},\phi^{m}) \times \right. \nonumber \\
    &&\left. \times \log (P(I_0(k)=\hat{y}_{0_k}, e_k^i=j, d_{k, j}^i=l\vert V_0(k)=\hat{y}_{0_k}, \hat{B}(k-1)=\hat{b}(k-1); p,\phi))\right] \nonumber \\
    &&+\sum_{k=1}^{T}\sum_{j=1}^{N}\sum_{l=1}^{j}\left[P(e_k^i=j \vert I_0(k)=\hat{z}_{0_k},V_0(k)=\hat{y}_{0_k}, \hat{B}(k-1)=\hat{b}(k-1); p^{m},\phi^{m})\times \right. \nonumber \\
    &&\left. \times P(d_{k,j}^i=l \vert e_k^i = j,I_0(k)=\hat{z}_{0_k},V_0(k)=\hat{y}_{0_k}, \hat{B}(k-1)=\hat{b}(k-1);p^{m},\phi^{m}) \times \right. \nonumber \\
    &&\left. \times \log (P(not(V_0(k)\vee I_0(k))=\hat{u}_{0_k}, e_k^i=j,d_{k,j}^i=l \vert I_0(k)=z_{0_k},V_0(k)=y_{0_k}, \hat{B}(k-1)=\hat{b}(k-1); p,\phi))\right]\nonumber \\
    \label{sec4.sec6.rem1.eq1}
\end{eqnarray}
where $\Theta = (p,\phi)$, and $(\hat{p}^{m},\hat{\phi}^{m})$ is the estimate of $(p,\phi)$ in the $m^{th}$ step of the EM algorithm.
\end{rem}
We specify an explicit expression for components of the E-step (\ref{sec4.sec6.rem1.eq1}) in the following result.
\begin{lemma}\label{sec4.sec6.lemma2}
For each $k \in \{1,2,3,\dots,T \}$, $j \in \{1,2,3,\dots,N\}$, and $l \in \{1,2,3,\dots,j\}$, the following holds:
    \begin{equation}\label{sec4.sec6.lemma2.eq1}
        P(e_k^i=j\vert \hat{B}(k-1)=\hat{b}(k-1); \hat{p}^{(m)},\hat{\phi}^{(m)})=\binom{N}{j}\left( \frac{\hat{I}(k-1)}{n-1}\right)^j\left(1-\frac{\hat{I}(k-1)}{n-1}\right)^{N-j},
    \end{equation}
        \begin{equation}\label{sec4.sec6.lemma2.eq2}
        P(e_k^i=j\vert V_{0}(k)=y_{0_{k}}, \hat{B}(k-1)=\hat{b}(k-1); \hat{p}^{(m)},\hat{\phi}^{(m)})=\binom{N}{j}\left( \frac{\hat{I}(k-1)}{n-1}\right)^j\left(1-\frac{\hat{I}(k-1)}{n-1}\right)^{N-j},
    \end{equation}
    and
    \begin{equation}\label{sec4.sec6.lemma2.eq2.eq1}
      P(e_k^i=j \vert I_0(k)=\hat{z}_{0_k},V_0(k)=\hat{y}_{0_k}, \hat{B}(k-1)=\hat{b}(k-1); \hat{p}^{(m)},\hat{\phi}^{(m)})\binom{N}{j}\left( \frac{\hat{I}(k-1)}{n-1}\right)^j\left(1-\frac{\hat{I}(k-1)}{n-1}\right)^{N-j}.
    \end{equation}
    Also,
    \begin{equation}\label{sec4.sec6.lemma2.eq3}
        P(d_{k,j}^i=l\vert e_k^i=j, \hat{B}(k-1)=\hat{b}(k-1);\hat{p}^{(m)},\hat{\phi}^{(m)})=\hat{p}^{(m)},
    \end{equation}
        \begin{equation}\label{sec4.sec6.lemma2.eq4}
        P(d_{k,j}^i=l\vert e_k^i=j, V_{0}(k)=y_{0_{k}}, \hat{B}(k-1)=\hat{b}(k-1);\hat{p}^{(m)},\hat{\phi}^{(m)})=\hat{p}^{(m)},
    \end{equation}
    and
            \begin{equation}\label{sec4.sec6.lemma2.eq4.eq1}
        P(d_{k,j}^i=l\vert e_k^i=j, I_{0}(k)=z_{0_{k}}, V_{0}(k)=y_{0_{k}}, \hat{B}(k-1)=\hat{b}(k-1);\hat{p}^{(m)},\hat{\phi}^{(m)})=\hat{p}^{(m)},
    \end{equation}
    Furthermore,
    \begin{eqnarray}
        P(V_0(k)=\hat{y}_{0_k}, e_k^i=j, d_{k,j}^i=l \vert \hat{B}(k-1)=\hat{b}(k-1); p,\phi) =\nonumber \\
        \binom{\underbrace{\hat{x}_{k-1}}}{\hat{y}_{0_k}}\binom{N}{j}\left(\frac{\hat{I}(k-1)}{n-1}\right)^j\left( 1- \frac{\hat{I}(k-1)}{n-1}\right)^{N-j}(\phi(1-p))^{y_{0_k}}(1-\phi(1-p))^{\underbrace{\hat{x}_{k-1}}-\hat{y}_{0_k}}p,\nonumber\\
        \label{sec4.sec6.lemma2.eq5}
    \end{eqnarray}
    \begin{eqnarray}
        P(I_0(k)=\hat{z}_{0_k}, e_k^i=j, d_{k,j}^i=l \vert V_0(k)=\hat{y}_{0_k}, \hat{B}(k-1)=\hat{b}(k-1); p,\phi) =\nonumber \\
        \binom{\hat{x}_{k-1}-\hat{y}_{0_k}}{\hat{z}_{0_k}}\binom{N}{j}\left(\frac{\hat{I}(k-1)}{n-1}\right)^j\left( 1- \frac{\hat{I}(k-1)}{n-1}\right)^{N-j}p^{(\hat{z}_{0_k}+1)}(1-p)^{{\hat{x}_{k-1}}-\hat{y}_{0_k}-\hat{z}_{0_k}},\label{sec4.sec6.lemma2.eq6}
    \end{eqnarray}
    and
   \begin{eqnarray}
        P(not(V_0(k)\vee I_0(k))=\hat{u}_{0_k}, e_k^i=j, d_{k,j}^i=l \vert V_0(k) = \hat{y}_{0_k}, \hat{B}(k-1)=\hat{b}(k-1); p,\phi)  =\nonumber \\
        (1)\binom{N}{j}\left(\frac{\hat{I}(k-1)}{n-1}\right)^j\left( 1- \frac{\hat{I}(k-1)}{n-1}\right)^{N-j}((1-p)(1-\phi))^{\hat{u}_{0_k}}p,\label{sec4.sec6.lemma2.eq6}
    \end{eqnarray}
    where $\hat{u}_{0_k}=\hat{x}_{k-1}-\hat{y}_{0_k}-\hat{z}_{0_k},\forall k$.
\end{lemma}
Proof:\\
The equations (\ref{sec4.sec6.lemma2.eq1})-(\ref{sec4.sec6.lemma2.eq4.eq1}) follow immediately from assumptions (a)-(d) in Subsection~\ref{sec4.subsec1}.  For (\ref{sec4.sec6.lemma2.eq5}) and (\ref{sec4.sec6.lemma2.eq6}), we apply the multiplication rule and also apply  assumptions (a)-(d) in Subsection~\ref{sec4.subsec1}.  That is,
\begin{eqnarray}
&&P(V_0(k) = \hat{y}_{0_k}, e_k^i=j, d_{k,j}^i=l \vert \hat{B}(k-1)=\hat{b}(k-1); p,\phi) =\nonumber \\
&&P(V_0(k) = \hat{y}_{0_k}, \vert e_k^i=j, d_{k,j}^i=l \hat{B}(k-1)=\hat{b}(k-1); p,\phi)\times \nonumber \\
&&\times P(d_{k,j}^i=l \vert e_k^i=j,  \hat{B}(k-1)=\hat{b}(k-1); p,\phi) \times \nonumber \\
&&\times P(e_k^i=j \vert \hat{B}(k-1)=\hat{b}(k-1); p,\phi).\label{sec4.sec6.lemma2.proof.eq1}
\end{eqnarray}
Observe from assumptions (a)-(d) in Subsection~\ref{sec4.subsec1} that, given infection is passed across only by the $l^{th}$  infectious person among the $j$ infectious persons encountered for that given instant, and also given that $\phi$ is the probability that a susceptible person changes the mind, and becomes vaccinated within that time instant after  escaping infection from the $l^{th}$  infectious person, it follows that $\phi (1-p)$ is the probability that a susceptible person gets vaccinated at any instant, and the conditional distribution of the random variable $V_0(k)$ is binomial with parameters $\phi (1-p)$ and $\underbrace{\hat{x}_{k-1}}$. Thus,
\begin{equation}\label{sec4.sec6.lemma2.proof.eq2}
    P(V_0(k) = \hat{y}_{0_k}\vert e_k^i=j, d_{k,j}^i=l, \hat{B}(k-1)=\hat{b}(k-1); p,\phi) = \binom{\underbrace{\hat{x}_{k-1}}}{\hat{y}_{0_k}}(\phi(1-p))^{y_{0_k}}(1-\phi(1-p))^{\underbrace{\hat{x}_{k-1}}-\hat{y}_{0_k}},
\end{equation}
Also,  the probability that the $l^{th}$ infectious person passes infection is given by
\begin{equation}\label{sec4.sec6.lemma2.proof.eq3}
    P(d_{k,j}^i=l \vert e_k^i=j, \hat{B}(k-1)=\hat{b}(k-1); p,\phi)=p,
\end{equation}
and the probability that the $i^{th}$ susceptible person meets $j$ infectious people at time $k$ is given by
\begin{equation}\label{sec4.sec6.lemma2.proof.eq4}
    P(e_k^i=j \vert \hat{B}(k-1)=\hat{b}(k-1); p,\phi)=\binom{N}{j}\left( \frac{\hat{I}(k-1)}{n-1}\right)^j \left( 1- \frac{\hat{I}(k-1)}{n-1}\right)^{N-j}.
\end{equation}
Substituting (\ref{sec4.sec6.lemma2.proof.eq2})-(\ref{sec4.sec6.lemma2.proof.eq4}) into (\ref{sec4.sec6.lemma2.proof.eq1}) gives (\ref{sec4.sec6.lemma2.eq5}).

Similarly,
 \begin{eqnarray}
&&P(I_0(k) = \hat{z}_{0_k}, e_k^i=j, d_{k,j}^i=l \vert V_0(k) = \hat{y}_{0_k}, \hat{B}(k-1)=\hat{b}(k-1); p,\phi) =\nonumber \\
&&P(I_0(k) = \hat{z}_{0_k}, \vert e_k^i=j, d_{k,j}^i=l, V_0(k) = \hat{y}_{0_k}, \hat{B}(k-1)=\hat{b}(k-1); p,\phi)\times \nonumber \\
&&\times P(d_{k,j}^i=l \vert e_k^i=j, V_0(k) = \hat{y}_{0_k}, \hat{B}(k-1)=\hat{b}(k-1); p,\phi) \times \nonumber \\
&&\times P(e_k^i=j \vert V_0(k) = \hat{y}_{0_k}, \hat{B}(k-1)=\hat{b}(k-1); p,\phi).\label{sec4.sec6.lemma2.proof.eq5}
\end{eqnarray}
From assumptions (a)-(d) in Subsection~\ref{sec4.subsec1}, it is easy to see that
\begin{equation}\label{sec4.sec6.lemma2.proof.eq6}
    P(I_0(k) = \hat{z}_{0_k}\vert  e_k^i=j, d_{k,j}^i=l, V_0(k) = \hat{y}_{0_k},\hat{B}(k-1)=\hat{b}(k-1); p) = \binom{\hat{x}_{k-1}-\hat{y}_{0_k}}{\hat{z}_{0_k}}(1-p)^{\hat{x}_{k-1}-\hat{y}_{0_k}-\hat{z}_{0_k}}(p)^{\hat{z}_{0_k}}.
\end{equation}
Furthermore, all the other components of (\ref{sec4.sec6.lemma2.proof.eq5}) are obtained similarly as in (\ref{sec4.sec6.lemma2.proof.eq3})-(\ref{sec4.sec6.lemma2.proof.eq4}).

Finally,
 \begin{eqnarray}
&&P(not(V_0(k)\vee I_0(k))=\hat{u}_{0_k}, e_k^i=j, d_{k,j}^i=l \vert V_0(k) = \hat{y}_{0_k}, \hat{B}(k-1)=\hat{b}(k-1); p,\phi) =\nonumber \\
&&P(not(V_0(k)\vee I_0(k))=\hat{u}_{0_k}, \vert e_k^i=j, d_{k,j}^i=l,I_0(k) = \hat{z}_{0_k}, V_0(k) = \hat{y}_{0_k}, \hat{B}(k-1)=\hat{b}(k-1); p,\phi)\times \nonumber \\
&&\times P(d_{k,j}^i=l \vert e_k^i=j,I_0(k) = \hat{z}_{0_k}, V_0(k) = \hat{y}_{0_k}, \hat{B}(k-1)=\hat{b}(k-1); p,\phi) \times \nonumber \\
&&\times P(e_k^i=j \vert I_0(k) = \hat{z}_{0_k},V_0(k) = \hat{y}_{0_k}, \hat{B}(k-1)=\hat{b}(k-1); p,\phi).\label{sec4.sec6.lemma2.proof.eq7}
\end{eqnarray}
From assumptions (a)-(d) in Subsection~\ref{sec4.subsec1}, it is easy to see that for $\hat{u}_{0_k}=\hat{x}_{k-1}-\hat{y}_{0_k}-\hat{z}_{0_k}$,
\begin{eqnarray}
   && P(not(V_0(k)\vee I_0(k))=\hat{u}_{0_k}\vert  e_k^i=j, d_{k,j}^i=l,I_0(k) = \hat{z}_{0_k}, V_0(k) = \hat{y}_{0_k},\hat{B}(k-1)=\hat{b}(k-1); p)\nonumber\\
    &&= \binom{\hat{x}_{k-1}-\hat{y}_{0_k}-\hat{z}_{0_k}}{\hat{u}_{0_k}}(p+\phi(1-p))^{\hat{x}_{k-1}-\hat{y}_{0_k}-\hat{z}_{0_k}-\hat{u}_{0_k}}(1-p-\phi(1-p))^{\hat{u}_{0_k}}\nonumber\\
    &&=(1-p-\phi(1-p))^{\hat{u}_{0_k}}.\label{sec4.sec6.lemma2.proof.eq8}
\end{eqnarray}
Furthermore, all the other components of (\ref{sec4.sec6.lemma2.proof.eq7}) are obtained similarly as in (\ref{sec4.sec6.lemma2.proof.eq3})-(\ref{sec4.sec6.lemma2.proof.eq4}).
Thus, from (\ref{sec4.sec6.lemma2.proof.eq5})-(\ref{sec4.sec6.lemma2.proof.eq8}), the result in (\ref{sec4.sec6.lemma2.eq6}) follows immediately.

The following result presents an expression for the E-step of the EM algorithm.
\begin{thm}\label{sec4.sec6.thm1}
    Assume that the results in Lemma~\ref{sec4.sec6.lemma2} hold. For $m=0, 1, 2, \dots$ , the E-step of the EM algorithm in (\ref{sec4.sec6.rem1.eq1}) in Remark~\ref{sec4.sec6.rem1} is expressed as follows for $\Theta=(p)$
    \begin{eqnarray}
        Q(\Theta \vert \hat{\Theta}^{(m)}) &\equiv&  \mathfrak{K} + \sum_{k=1}^{T}N\left(\frac{\hat{I}(k-1)}{n-1}\right)\hat{p}^{(m)}\times\nonumber\\
        &&\times\left([1+(\hat{z}_{0_k}+1)+1]\log{(p)}+(\hat{x}_{k-1}-\hat{z}_{0_k}+\hat{u}_{0_k})\log{(1-p)}\right.\nonumber\\
        &&\left. (\underbrace{\hat{x}_{k-1}}-\hat{y}_{0_k})\log(1-\phi(1-p))+\hat{y}_{0_k}\log(\phi)+\hat{u}_{0_k}\log(1-\phi)\right),\label{sec4.sec6.thm1.eq1}
                    \end{eqnarray}
where $\hat{u}_{0_k}=\hat{x}_{k-1}-\hat{y}_{0_k}-\hat{z}_{0_k},\forall k$, $\mathfrak{K}$ denotes a constant term, and $\hat{p}^{(m)}$ is an estimate of $p$ at the $m^{th}$ step. Also
\begin{equation}\label{sec4.sec6.thm1.eq2}
\hat{I}(k-1)=\hat{z}_{0_0}+\hat{z}_{0_1}+\cdots+\hat{z}_{0_{k-1}}.
\end{equation}
\end{thm}
Proof:\\
From Lemma~\ref{sec4.sec6.lemma1} and (\ref{sec4.sec6.rem1.eq1}), it is easy to see that

\begin{eqnarray}
        Q(\Theta \vert \hat{\Theta}^{(m)}) &\equiv& Q(p \vert \hat{p}^{(m)}) \nonumber \\
        &\equiv&\sum_{k=1}^{T}\sum_{j=1}^{N}\sum_{l=1}^{j}\binom{N}{j}\left( \frac{\hat{I}(k-1)}{n-1}\right)^j\left( 1-\frac{\hat{I}(k-1)}{n-1}\right)^{N-j}\hat{p}^{(m)}\times\nonumber \\
        &&\left[ \log  \left[\binom{\underbrace{\hat{x}_{k-1}}}{\hat{y}_{0_k-1}}\binom{N}{j}\left(\frac{\hat{I}(k-1)}{n-1}\right)^j\left(1-\frac{\hat{I}(k-1)}{n-1}\right)^{N-j}\right]+\right. \nonumber \\
        &&\left[ (\underbrace{\hat{x}_{k-1}} -\hat{y}_{0_k})\log (1-\phi(1-p))+(\hat{y}_{0_k})\log (1-p)+(\hat{y}_{0_k})\log (\phi)+\log(p)\right] \nonumber \\
        &&+\sum_{k=1}^{T}\sum_{j=1}^{N}\sum_{l=1}^{j}\binom{N}{j}\left( \frac{\hat{I}(k-1)}{n-1}\right)^j\left( 1-\frac{\hat{I}(k-1)}{n-1}\right)^{N-j}\hat{p}^{(m)}\times\nonumber \\
        &&\left[ \log  \left[\binom{\hat{x}_{k-1}-\hat{y}_{0_k}}{\hat{z}_{0_k-1}}\binom{N}{j}\left(\frac{\hat{I}(k-1)}{n-1}\right)^j\left(1-\frac{\hat{I}(k-1)}{n-1}\right)^{N-j}\right]+\right. \nonumber \\
        &&\left[ (\hat{z}_{0_k}+1)\log (p)+(\hat{x}_{k-1}-\hat{y}_{0_k}-\hat{z}_{0_k})\log (1-p)\right] \nonumber \\
        &&+\sum_{k=1}^{T}\sum_{j=1}^{N}\sum_{l=1}^{j}\binom{N}{j}\left( \frac{\hat{I}(k-1)}{n-1}\right)^j\left( 1-\frac{\hat{I}(k-1)}{n-1}\right)^{N-j}\hat{p}^{(m)}\times\nonumber \\
        &&\left[ \log  \left[\binom{N}{j}\left(\frac{\hat{I}(k-1)}{n-1}\right)^j\left(1-\frac{\hat{I}(k-1)}{n-1}\right)^{N-j}\right]+\right. \nonumber \\
        &&\left[ (\hat{u}_{0_k})\log (1-p)+(\hat{u}_{0_k})\log(1-\phi)+\log (p)\right]\nonumber \\\label{sec4.sec6.thm1.proof.eq1}
        %
%
%
\end{eqnarray}
where $\hat{u}_{0_k}=\hat{x}_{k-1}-\hat{y}_{0_k}-\hat{z}_{0_k},\forall k$.
Observe that
\begin{equation}\label{sec4.sec6.thm1.proof.eq2}
\sum_{j=1}^{N}\sum_{l=1}^{j}\binom{N}{j}\left( \frac{\hat{I}(k-1)}{n-1}\right)^j\left( 1-\frac{\hat{I}(k-1)}{n-1}\right)^{N-j}=N\left( \frac{\hat{I}(k-1)}{n-1}\right).
\end{equation}
Thus, (\ref{sec4.sec6.thm1.eq1}) follows immediately from (\ref{sec4.sec6.thm1.proof.eq1}), where
\begin{eqnarray}
        \mathfrak{K}
        &\equiv&\sum_{k=1}^{T}\sum_{j=1}^{N}\sum_{l=1}^{j}\binom{N}{j}\left( \frac{\hat{I}(k-1)}{n-1}\right)^j\left( 1-\frac{\hat{I}(k-1)}{n-1}\right)^{N-j}\hat{p}^{(m)}\times\nonumber \\
        &&\left[ \log  \left[\binom{\underbrace{\hat{x}_{k-1}}}{\hat{y}_{0_k-1}}\binom{N}{j}\left(\frac{\hat{I}(k-1)}{n-1}\right)^j\left(1-\frac{\hat{I}(k-1)}{n-1}\right)^{N-j}\right]\right] \nonumber \\
        &&+\sum_{k=1}^{T}\sum_{j=1}^{N}\sum_{l=1}^{j}\binom{N}{j}\left( \frac{\hat{I}(k-1)}{n-1}\right)^j\left( 1-\frac{\hat{I}(k-1)}{n-1}\right)^{N-j}\hat{p}^{(m)}\times\nonumber \\
        &&\left[ \log  \left[\binom{\hat{x}_{k-1}-\hat{y}_{0_k}}{\hat{z}_{0_k-1}}\binom{N}{j}\left(\frac{\hat{I}(k-1)}{n-1}\right)^j\left(1-\frac{\hat{I}(k-1)}{n-1}\right)^{N-j}\right]\right] \nonumber \\
        &&+\sum_{k=1}^{T}\sum_{j=1}^{N}\sum_{l=1}^{j}\binom{N}{j}\left( \frac{\hat{I}(k-1)}{n-1}\right)^j\left( 1-\frac{\hat{I}(k-1)}{n-1}\right)^{N-j}\hat{p}^{(m)}\times\nonumber \\
        &&\left[ \log  \left[\binom{N}{j}\left(\frac{\hat{I}(k-1)}{n-1}\right)^j\left(1-\frac{\hat{I}(k-1)}{n-1}\right)^{N-j}\right]\right]. \nonumber \\\label{sec4.sec6.thm1.proof.eq3}
        %
%
%
\end{eqnarray}
Also, (\ref{sec4.sec6.thm1.eq2}) follows from Definition~\ref{ch2.sec2.def4}.
\begin{rem}\label{sec4.sec6.rem2}
It follows from Theorem~\ref{sec4.sec6.thm1} that the M-step of the EM algorithm consists of maximizing $Q(p,\phi\vert \hat{p}^{(m)},\hat{\phi}^{(m)})$ with respect to $p,\phi$.  This is equivalent to maximizing the non-constant term of (\ref{sec4.sec6.thm1.eq1}).
\end{rem}
In the next result, we present the EM algorithm MLE estimator for $p,\phi$.
\begin{thm}\label{sec4.sec6.thm2}
    Let the E step of the EM algorithm be as defined in Theorem \ref{sec4.sec6.thm1}.  It follows that the MLE for $p$ is given as follows:
    \begin{eqnarray}
        \hat{p} &=& \frac{\sum_{k=1}^{T}N\left( \frac{\sum^{k-1}_{h=0}\hat{z}_{0_h}}{n-1}\right)(\underbrace{\hat{x}_{k-1}}+1-\hat{y}_{0_k}+\hat{z}_{0_k}+1)}{\sum_{k=1}^{T}N\left( \frac{\sum^{k-1}_{h=0}\hat{z}_{0_h}}{n-1}\right)(\underbrace{\hat{x}_{k-1}}+1-\hat{y}_{0_k}+\hat{x}_{k-1}+1)} \nonumber\\
        &=& \frac{\sum_{k=1}^{T}\left( \frac{\sum^{k-1}_{h=0}\hat{z}_{0_h}}{n-1}\right)(\underbrace{\hat{x}_{k-1}}+1-\hat{y}_{0_k}+\hat{z}_{0_k}+1)}{\sum_{k=1}^{T}\left( \frac{\sum^{k-1}_{h=0}\hat{z}_{0_h}}{n-1}\right)(\underbrace{\hat{x}_{k-1}}+1-\hat{y}_{0_k}+\hat{x}_{k-1}+1)}.\label{sec4.sec6.thm2.eq1}
    \end{eqnarray}
\end{thm}
Proof:\\
From (\ref{sec4.sec6.thm1.eq1}), it follows that the derivative
\begin{eqnarray}
\frac{d Q(\Theta \vert \hat{\Theta}^{(m)})}{d p} = \sum_{k=1}^{T}N\left( \frac{\hat{I}(k-1)}{n-1}\right)\hat{p}^{(m)}\left[[(\underbrace{\hat{x}_{k-1}}+1-\hat{y}_{0_k})+(\hat{z}_{0_k}+1)]\frac{1}{p}+(\hat{x}_{k-1}-\hat{z}_{0_k})(-1)\frac{1}{1-p}\right].\nonumber\\
\label{sec4.sec6.thm2.proof.eq1}
\end{eqnarray}
It follows from (\ref{sec4.sec6.thm2.proof.eq1}) that since $\hat{p}^{(m)} > 0$, then setting the right-hand-side to zero and solving for $p$ leads to
\begin{equation}\label{sec4.sec6.thm2.proof.eq2}
        \hat{p} = \frac{\sum_{k=1}^{T}N\left( \frac{\sum^{k-1}_{h=0}\hat{z}_{0_h}}{n-1}\right)(\underbrace{\hat{x}_{k-1}}+1-\hat{y}_{0_k}+\hat{z}_{0_k}+1)}{\sum_{k=1}^{T}N\left( \frac{\sum^{k-1}_{h=0}\hat{z}_{0_h}}{n-1}\right)(\underbrace{\hat{x}_{k-1}}+1-\hat{y}_{0_k}+\hat{x}_{k-1}+1)}.
    \end{equation}
Since for all $m=0,1,2,\dots$ the solution of $ \frac{d Q(\Theta \vert \hat{\Theta}^{(m)})}{d p}  =0$ from (\ref{sec4.sec6.thm2.proof.eq1}) remains the same and is given by (\ref{sec4.sec6.thm2.proof.eq2}), therefore the EM-algorithm converges for the value of $p$ in (\ref{sec4.sec6.thm2.eq1}).
\section{Some epidemiological parameters for evaluating the occurrence of epidemics}\label{sec5}
In this section, we calculate some epidemiological parameters to evaluate the prevalence of influenza.  We consider two disease control parameters, namely: the basic reproduction number, and the probability of no spread.  These parameters are used in  \cite{Impact, BRN, burgers, BRNnotes, BRNpredictor}.

Furthermore, these epidemiological parameters are calculated for the random process $\{B(t), t=0, 1, 2, \dots\}$ whenever the transition probabilities are defined in Theorem~\ref{sec4.subsec1.thm1}.  

In the following, we characterize the prevalence of influenza from the initial infected population.  That is, we calculate the expected number of infected individuals that occur over time, given an initial infected population.  This information is useful to determine whether an epidemic will occur from the initial infected population.
\subsection{Expected number of infected individuals}\label{sec5.subsec1}
Recall Definition~\ref{ch2.sec2.def4}, for $t<min\{T_1, T_2\}$, and $T_1 \leq t < T_2$,
\begin{equation}\label{A01}
I(t) = I_0(t)+ I_1(t)+ I_2(t) + \dots+ I_t(t).
\end{equation}
Furthermore, for $T_2 \leq t < T_1$ and $t \geq max\{T_1, T_2\}$,
\begin{equation}\label{A02}
I(t) = I_0(t)+ I_1(t)+ \dots+ I_{T_2-2}+ I_{T_2-1}.
\end{equation}

In the following result, we show that the expected infectious population over time, given the initial outbreak of influenza depends only the state of the process at one time lag.
\begin{lemma}\label{th6.1}
For any $t=0,1,2,\dots$,
\begin{equation}\label{th6.1.eq1}
E[I_0(t)\vert B(t-1), B(t-2), \dots, B(1), B(0)]=E[I_0(t)\vert B(t-1)],
\end{equation}
and
\begin{equation}\label{A03}
E[I_0(t)\vert B(0)] = E[E[I_0(t)\vert B(t-1)]\vert B(0)] .
\end{equation}
\end{lemma}
Proof:\\
It is easy to see from Theorem~\ref{ch2.sec3.thm1} and Theorem~\ref{sec4.subsec1.thm1} that
\begin{eqnarray}
&&E[I_0(t)\vert B(t-1), B(t-2), \dots, B(1), B(0)]\nonumber\\
&&=\sum_{z}\sum_{y}zP(V_0(t)=y, I_0(t)=z\vert B(t-1), B(t-2), \dots, B(1), B(0))\nonumber\\
&&=\sum_{z}\sum_{y}zP(V_0(t)=y, I_0(t)=z\vert B(t-1))\nonumber\\
&&= E[I_0(t)\vert B(t-1)].\label{th6.1.proof.eq1}
\end{eqnarray}
Also, by the properties of conditional expectations, for any $t=0,1,2,\ldots$,
\begin{equation}\label{A04}
E[I_0(t)\vert B(0)] = E[E[I_0(t)\vert B(t-1), B(t-2), \dots, B(1), B(0)]\vert B(0)] .
\end{equation}
From (\ref{th6.1.proof.eq1}), it follows that (\ref{A04}) reduces to,
\begin{equation}\label{A041}
E[I_0(t)\vert B(0)] = E[E[I_0(t)\vert B(t-1)]\vert B(0)]
\end{equation}
definitely.

Using Lemma~\ref{th6.1}, we present in general form the expected number of infectious people present at any time $t$, given the population at the initial outbreak.
\begin{thm}\label{Th6.2}
Let the assumptions of Theorem~\ref{ch2.sec3.thm1} and Theorem~\ref{sec4.subsec1.thm1}  hold. For $t<min\{T_1,T_2\}$ and $t \in [T_1, T_2)$, it follows that
\begin{equation}\label{A05}
E[I(t)\vert B(0)] = I_0(0) + \sum_{k=1}^{t-1}E\left[S(t-k-1)\left(1-\left(1-\frac{pI(t-k-1)}{n-1}\right)^N\right)\vert B(0)\right].
\end{equation}
For $t \in [T_2,T_1)$ and $t \geq max \{T_1, T_2\}$,
\begin{equation}\label{A06}
E[I(t)\vert B(0)] = \sum_{k=1}^{T_2-1} E\left[S(t-k-1)\left(1-\left(1-\frac{pI(t-k-1)}{n-1}\right)^N\right)\vert B(0)\right].
\end{equation}
\end{thm}
Proof:\\
For $t < min \{T_1,T_2\}$,
\begin{eqnarray}\label{eq6.2}
I(t) &=& I_0(t) + I_1(t) + I_2(t) + \dots + I_t(t) \nonumber\\
&=& I_0(t)+I_0(t-1) +I_0(t-2) + \dots + I_0(1)+ I_0(0) .
\end{eqnarray}
\begin{eqnarray}\label{eq6.4}
E[I(t)\vert B(0)] &=& E[I_0(t)\vert B(0)]+E[I_0(t-1)\vert B(0)] + \dots \nonumber \\
&& + E[I_0(1)\vert B(0)] + E[I_0(0)\vert B(0)] \nonumber\\
&=& I_0(0)+ \sum_{k=1}^{t-1} E[I_0(t-k)\vert B(0)] .
\end{eqnarray}
For each $ k=1,2,3,\ldots,t-1$,  applying Lemma~\ref{th6.1}
\begin{eqnarray}\label{eq6.8}
E[I_0(t-k)\vert B(0)] &=& E[E[I_0(t-k)\vert B(t-k-1)]\vert B(0)]\nonumber\\
&=& E[S(t-k-1)P_I(t-k)\vert B(0)]\nonumber\\
&=& E\left[S(t-k-1)\left(1-\left(1- \frac{pI(t-k-1)}{n-1}\right)^N \right)\vert B(0)\right],\nonumber \\
\end{eqnarray}
where $P_I(t-k)$ is defined in (\ref{4.2C}).  Substituting (\ref{eq6.8}) into (\ref{eq6.4}), we obtain the result in (\ref{A05}).  Observe that the result for $t \in [T_1, T_2)$ is obtained similarly as above.

For $t \in [T_2,T_1)$ and $t > max\{T_1, T_2\}$
\begin{eqnarray}\label{eq6.20}
I(t) &=& I_0(t) + I_1(t) + I_2(t) + \dots + I_{T_2-1}(t) \nonumber \\
&=& I_0(t)+I_0(t-1) +I_0(t-2) + \dots +I_0(t-(T_2-1)) .
\end{eqnarray}
\begin{eqnarray}\label{eq6.22}
E[I(t)\vert B(0)] &=& E[I_0(t)\vert B(0)]+E[I_0(t-1)\vert B(0)] + \dots \nonumber \\
&& + E[I_0(t-(T_2-1)\vert B(0)]\nonumber \\
&=& \sum_{k=0}^{T_2-1} E[I_0(t-k)\vert B(0)] .
\end{eqnarray}
For each $ k=0,1,2,3,\dots,(T_2-1)$  applying Lemma~\ref{th6.1}\\
\begin{eqnarray}\label{eq6.26}
E[I_0(t-k)\vert B(0)] &=& E[E[I_0(t-k)\vert B(t-k-1)]\vert B(0)]\nonumber\\
&=& E[S(t-k-1)P_I(t-k-1)\vert B(0)]\nonumber\\
&=& E\left[S(t-k-1)\left(1-\left(1- \frac{pI(t-k-1)}{n-1}\right)^N \right)\vert B(0)\right],\nonumber \\
\end{eqnarray}
where $P_I$ is defined in (\ref{4.2C}).  Substituting (\ref{eq6.26}) into (\ref{eq6.22}), we obtain (\ref{A06}).
\begin{rem}\label{sec5.subsec1.thm1.rem1}
It should be observed from Theorem~\ref{Th6.2} that an explicit form for the conditional expectation (\ref{A05}) and (\ref{A06}) can only be obtained provided the joint distribution of $(S(t), I(t)), \forall t \geq 0$ is known.  However, since (\ref{A05}) and (\ref{A06}) represent population parameters at time $t$ (conditional population means), which are a sum of random variables that represent observations over time until the time $t$, these parameters can be estimated point-wise using sample paths of the process $\{B(t), t=0, 1, 2, \dots\}$, and the MLE of $p$ obtained in Theorem~\ref{sec4.sec6.thm2}.

 For example, for $t > max \{T_1, T_2\}$, (\ref{A06}) can be estimated using the sample path of
\begin{eqnarray}
&&(S(t-k-1), I_0(t-k-1), I_1(t-k-1), \dots, I_{t-k-1}(t-k-1)) =\nonumber \\
&&(\hat{x}_{t-k-1}, \hat{z}_{0_{t-k-1}}, \hat{z}_{1_{t-k-1}}, \dots, \hat{z}_{{t-k-1}_{t-k-1}}),\nonumber \\
&&k=0,1,2,\dots, T_2-1.\label{sec5.subsec1.thm1.rem1.eq1}
\end{eqnarray}
\end{rem}
The next result presents the estimates for the conditional population means at any time $t$,
\begin{thm}\label{sec5.subsec1.thm2}
Assume that the conditions of Theorem~\ref{Th6.2} are satisfied.
 For  $t > max \{T_1, T_2\}$, the conditional expected value in (\ref{A06}) denoted  $\mu_{I(t)\vert B(0)}=E[I(t)\vert B(0)]$ can be estimated using the sample path of the process $\{ B(t),t=0,1,2,\dots \}$ namely:-
\begin{eqnarray}
&&(S(t-k-1), I_0(t-k-1), I_1(t-k-1), \dots, I_{t-k-1}(t-k-1)) =\nonumber \\
&&(\hat{x}_{t-k-1}, \hat{z}_{0_{t-k-1}}, \hat{z}_{1_{t-k-1}}, \dots, \hat{z}_{{t-k-1}_{t-k-1}}),\nonumber \\
&&k=0,1,2,\dots, T_2-1.\label{sec5.subsec1.thm2.eq1}
\end{eqnarray}
In fact, for $t> max\{T_1, T_2\}$ and $t\in [T_2, T_1)$
\begin{equation}\label{eq6.37}
    \hat{\mu}_{I(t)\vert B(0)}=\sum_{k=0}^{T_2-1}\left[x_{t-k-1}\left(1-\left(1-\frac{\hat{p}\sum_{j=0}^{t-k-1}z_{j_{t-k-1}}}{n-1}\right)^N\right)\right]
\end{equation}
estimates $\mu_{I(t)\vert B(0)}=E[I(t)\vert B(0)]$ defined in (\ref{A06}).

Also, for $t<min\{T_1, T_2\}$ and $t\in [T_1, T_2)$
\begin{equation}\label{eq6.38}
    \hat{\mu}_{I(t)\vert B(0)}=z_{0_0}+\sum_{k=1}^{t-1}\left[x_{t-k-1}\left(1-\left(1-\frac{\hat{p}\sum_{j=0}^{t-k-1}z_{j_{t-k-1}}}{n-1}\right)^N\right)\right]
\end{equation}
estimates $\mu_{I(t)\vert B(0)}=E[I(t)\vert B (0)]$ defined in (\ref{A05}), where $\hat{p}$ is the MLE of $p$ defined in (\ref{sec4.sec6.thm2.eq1}).
\end{thm}
Proof:\\
The results follow simply from (\ref{sec4.sec6.thm1.eq2}) and the definition of $\mu_{I(t)\vert B(0)}=E[I(t)\vert B (0)]$ defined in (\ref{A06}) and (\ref{A05}).
\subsection{The basic reproduction number for the SVIR influenza epidemic}
The basic reproduction number, generally denoted $R_0$, is the expected number of secondary cases of infection from one infectious person or from an initial number of infectious people, $I(0) = z_{0_0}$, placed in a completely susceptible population, and it is the most widely used predictor of an epidemic outbreak.  If $R_0< 1$, the disease is expected to die out.  If $R_0>1$, the disease is expected to spread out of control.

The basic reproduction number is highly dependent on the initial infectiousness of the disease and the duration of the disease in the initial infectious population.  The basic reproduction number is important to determine disease control factors for an epidemic \cite{BRN, burgers, BRNnotes, BRNpredictor}. In the absence of an explicit formula for the basic reproduction number $R_0$, it can be estimated empirically using the methods in this section.

 Observe from (\ref{A05}) that while $I_{0}(0)$ is the initial infectious population, the second term $\sum_{k=1}^{t-1}E\left[S(t-k-1)\left(1-\left(1-\frac{pI(t-k-1)}{n-1}\right)^N\right)\vert B(0)\right]$ represents the expected number of secondary infectious cases present at time $t$, given that the initial disease outbreak had only $I_{0}(0)$ number of infectious cases. Therefore, $\sum_{k=1}^{T_{2}-1}E\left[S(T_{2}-k-1)\left(1-\left(1-\frac{pI(T_{2}-k-1)}{n-1}\right)^N\right)\vert B(0)\right]$ must be the basic reproduction number given that $I_{0}(0)=1$, where $T_{2}$ is the constant infectious period of every infectious person in the population.
 Using the results of Theorem~\ref{sec5.subsec1.thm2},  the basic reproduction number $R_0$ can be estimated the following.

\begin{corollary}\label{sec5.subsec1.thm2.corol1}
Let the assumptions of Theorem~\ref{Th6.2} and Theorem~\ref{sec5.subsec1.thm2} be satisfied. The basic reproduction number is given implicitly as follows:
\begin{equation}\label{ch5.eq19}
    R_0=E[I(T_2)\vert B(0)]=\sum_{k=1}^{T_2-1}E\left[S\left(T_2-k-1\right)\left(1-\left(1-\frac{pI(T_2-k-1)}{n-1}\right)^N\right)\vert B(0)\right],
\end{equation}
%
For $t\leq T_2$, (\ref{ch5.eq19}) can be estimated using the following sample path of the process $\{B(t), t=0, 1, 2, \dots\}$
\begin{eqnarray}
&&\left( S(T_2-k-1), I_0(T_2-k-1), I_1(T_2-k-1), \dots, I_{T_2-k-1}(T_2-k-1)\right) =\nonumber \\
&&(x_{T_2-k-1}, z_{0_{T_2-k-1}}, z_{1_{T_2-k-1}}, \dots, z_{{T_2-k-1}_{T_2-k-1}}),\nonumber \\
&&k=1,2,\dots, T_2-1. \label{sec5.subsec1.thm2.corol1.eq1}
\end{eqnarray}
Furthermore, the estimate for $R_0$ is given as follows
\begin{equation}\label{sec5.subsec1.thm2.corol1.eq2}
    \hat{R}_0=\sum_{k=1}^{T_2-1}x_{T_{2}-k-1}\left(1-\left(1-\frac{\hat{p}\sum_{j=0}^{T_2-k-1}z_{j_{T_2-k-1}}}{n-1}\right)^N\right).
\end{equation}
\end{corollary}
Proof:\\
The result follows very easily from the definition of the basic reproduction number,  (\ref{eq6.38}), and setting $t=T_{2}$, since $T_{2}$ is the infectious period of individuals in the population.
\begin{rem}
It should be observed from Corollary~\ref{sec5.subsec1.thm2.corol1} that an explicit form for the basic reproduction number in (\ref{ch5.eq19}) can only be obtained provided the joint distribution of $(S(t), I(t)) \forall t \geq 0$ is known.  However, since (\ref{ch5.eq19}) represents a population parameter at time $t$, which is a sum of random variables that represent observations over time until the time $t$, this parameter is easily estimated-point using the sample path of the process $\{B(t), t=0, 1, 2, \dots\}$  given in (\ref{sec5.subsec1.thm2.corol1.eq1}), and the MLE of $p$ obtained in (\ref{sec4.sec6.thm2.proof.eq2}).

The advantage of  $\hat{R}_0$ in (\ref{sec5.subsec1.thm2.corol1.eq2}) as an initial point estimate for the actual value $R_0$   is its dependence primarily on empirical data for influenza obtained over time, and also the dependence on the feasible estimated value of the probability of passing infection from one infectious contact $p$. Furthermore, with limited real data over several possible sample realizations for the stochastic process $\{B(t), t=0, 1, 2, \dots\}$ over a given time interval, the statistic $\hat{R}_0$  can  be studied numerically and the approximate sampling distribution generated and studied.

More interval estimates for the parameter $R_0$ in (\ref{ch5.eq19}), require explicit sampling distribution for the point estimate  $\hat{R}_0$, which are beyond the scope of this work.
\end{rem}
\section{Conclusion}
In this study, we have sufficiently defined an SVIR Markov chain model for influenza which effectively shows the progression of the disease and effectiveness of the vaccine to control the disease over time for an individual in the population.  Moreover, we defined the transition probabilities for the model.  We presented two special cases of our model-(1)  based on the assumption that the  event of getting vaccinated at any instant depends on encounter with infectious people, and (2) the vaccination occurring over time with as a poisson process. We present detailed derivations of the probabilities of the $i^{th}$ susceptible individual getting vaccinated or infected at any instant, and further define the transition probabilities for each special case of the model.

We further used the maximum likelihood estimation technique and the expectation maximization (EM) algorithm to approximate the fixed parameters of the model.  To evaluate the occurrence and prevalence of the epidemic, we derived estimators for the basic reproduction number $R_0$, and the expected number of infected individuals at any time $t$.

Finally, we presented numerical simulation examples for the  influenza epidemic, and approximate the distribution of the total number of people in the population who ever get infected.  Given these scenarios, we see how vaccination can curb the spread of influenza, and that the most limiting factor when trying to control the spread of influenza is the number of infectious people one interacts with per unit time.

\end{document}